\documentclass[twocolumn,english,prl, twocolumns,superscriptaddress]{revtex4-1}
\usepackage[T1]{fontenc}
\usepackage[latin9]{inputenc}
\usepackage{verbatim}
\usepackage{float}
\usepackage{amsmath}
\usepackage{amssymb}
\usepackage{graphicx}
\usepackage{esint}
\usepackage{babel}
\usepackage{epstopdf}

\begin{document}
\global\long\def\dyad#1{\underline{\underline{\boldsymbol{#1}}}}
\global\long\def\ubar#1{\underbar{\ensuremath{\boldsymbol{#1}}}}
\global\long\def\Z{\mathbb{Z}}
\global\long\def\N{\mathbb{N}}
\global\long\def\R#1{\mathbb{R}^{#1}}
\global\long\def\C#1{\mathbb{C}^{#1}}
\global\long\def\defined{\triangleq}
\global\long\def\trace{\text{trace}}
\global\long\def\del{\nabla}
\global\long\def\cross{\times}
\global\long\def\diff#1#2{\frac{\partial#1}{\partial#2}}
\global\long\def\Diff#1#2{\frac{d#1}{d#2}}
\global\long\def\bra#1{\left\langle #1\right|}
\global\long\def\ket#1{\left|#1\right\rangle }
\global\long\def\braket#1#2{\left\langle #1|#2\right\rangle }
\global\long\def\ketbra#1#2{\left|#1\right\rangle \left\langle #2\right|}
\global\long\def\identity{\mathbf{1}}
\global\long\def\paulix{\begin{pmatrix}  &  1\\
1
\end{pmatrix}}
\global\long\def\pauliy{\begin{pmatrix}  &  -i\\
i
\end{pmatrix}}
\global\long\def\pauliz{\begin{pmatrix}1\\
 &  -1
\end{pmatrix}}
\global\long\def\sinc{\mbox{sinc}}
\global\long\def\four{\mathcal{F}}
\global\long\def\dag{^{\dagger}}
\global\long\def\norm#1{\left\Vert #1\right\Vert }
\global\long\def\hamil{\mathcal{H}}
\global\long\def\tens{\otimes}
\global\long\def\ord#1{\mathcal{O}\left(#1\right)}
\global\long\def\undercom#1#2{\underset{_{#2}}{\underbrace{#1}}}
 \global\long\def\conv#1#2{\underset{_{#1\rightarrow#2}}{\longrightarrow}}
\global\long\def\tg{^{\prime}}
\global\long\def\ttg{^{\prime\prime}}
\global\long\def\clop#1{\left[#1\right)}
\global\long\def\opcl#1{\left(#1\right]}
\global\long\def\broket#1#2#3{\bra{#1}#2\ket{#3}}
\global\long\def\div{\del\cdot}
\global\long\def\rot{\del\cross}

\title{Driving induced many-body localization}


\author{Eyal Bairey}
\affiliation{Physics Department, Technion, 3200003, Haifa, Israel}

\author{Gil Refael}
\affiliation{Institute for Quantum Information and Matter, Caltech, Pasadena,
California 91125, USA}
\author{Netanel H. Lindner}
\affiliation{Physics Department, Technion, 3200003, Haifa, Israel}

\begin{abstract}
Subjecting a many-body localized system to a time-periodic drive generically
leads to delocalization and a transition to ergodic behavior if the
drive is sufficiently strong or of sufficiently low frequency. Here
we show that a specific drive can have an opposite effect, taking
a static delocalized system into the many-body localized phase. We demonstrate this
effect using a one-dimensional system of interacting hardcore bosons
subject to an oscillating linear potential. The system is weakly disordered,
and is ergodic absent the driving. The time-periodic linear potential
leads to a suppression of the effective static hopping amplitude,
increasing the relative strengths of disorder and interactions. Using
numerical simulations, we find a transition into the many-body localized phase above
a critical driving frequency and in a range of driving amplitudes.
Our findings highlight the potential of driving schemes exploiting
the coherent destruction of tunneling for engineering long-lived
Floquet phases.
\end{abstract}
\maketitle
A key obstacle in the search for new non-equilibrium quantum phases
of matter is the tendency of closed quantum many-body systems to indefinitely
absorb energy from a time-periodic driving field. Thus, in the long
time limit, such systems generically reach a featureless infinite-temperature-like
state with no memory of their initial conditions \cite{DAlessio2014,Lazarides2014,Lazarides2014a,Chandran2016,Citro2015,Kukuljan2016,DAlessio2013,Ponte2015a}.
Interestingly, this infinite temperature fate can be avoided by the addition of disorder \cite{Lazarides2015,Ponte2015,Abanin2016,Rehn2016,Gopalakrishnan2016a}.
Sufficiently strong disorder added to a clean interacting system may
lead to a many-body localized (MBL) phase \cite{Anderson1958,Basko2006,Gornyi2005,Oganesyan2007,Nandkishore2015} which does not allow transport of energy and particles. The MBL phase
can persist in the presence of a weak, high-frequency drive \cite{Lazarides2015,Ponte2015,Abanin2016,Rehn2016,Gopalakrishnan2016a}.
Periodically driven systems in the MBL phase retain memory of their initial conditions for arbitrarily long times. Thus, they can support non-equilibrium
quantum phases of matter, including some which are unique to the non-equilibrium
setting \cite{Moessner2017,Titum2016,PhysRevLett.116.250401,VonKeyserlingk2016a,VonKeyserlingk2016b,VonKeyserlingk2016,Else2016,Else2016a,Potirniche2016,Potter2016,Magnetization,PhysRevX.6.041070}.

\begin{figure}[h!]
\includegraphics[width=8.6cm]{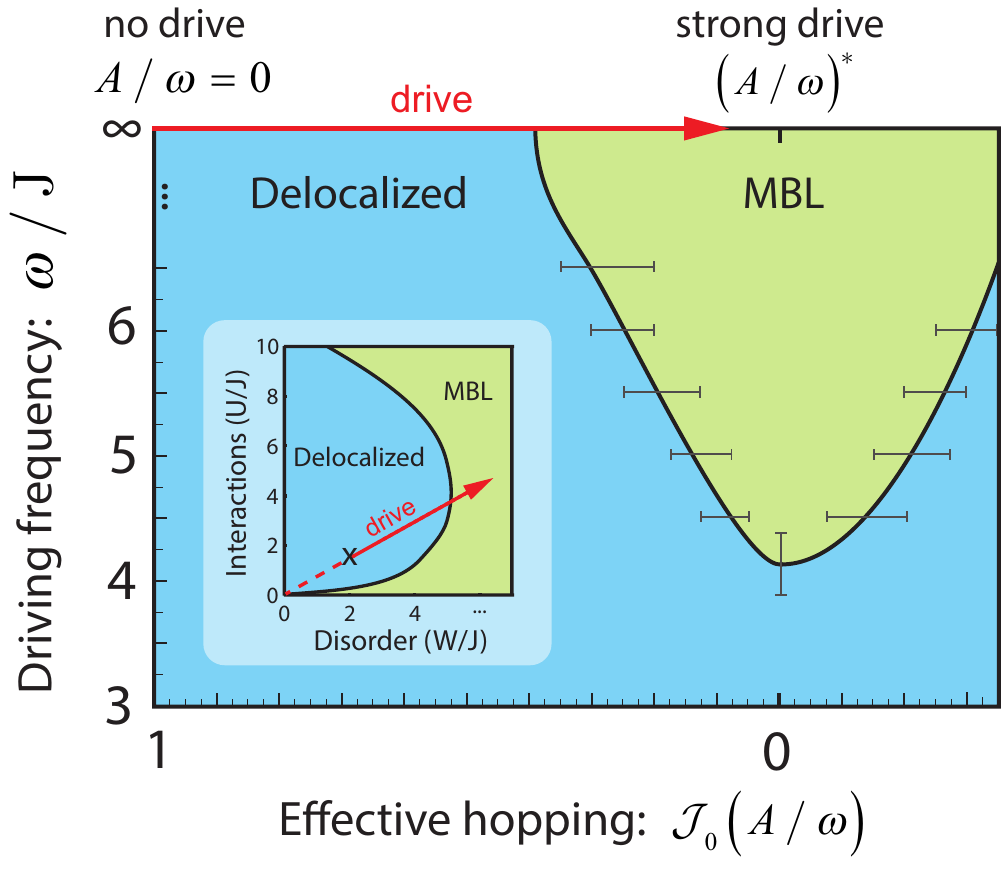}

\protect\caption{Phase diagram showing the MBL phase induced by time-periodic driving.
Hardcore bosons with nearest-neighbor interactions on a weakly disordered
1D lattice enter an MBL phase when driven by an AC electric field
above a critical driving frequency and in a range of driving amplitudes.
The horizontal axis corresponds to the AC field's amplitude $A$, measured by the effective hopping amplitude
$J_{eff}/J=\mathcal{J}_{0}\left(A/\omega\right)$ where $\mathcal{J}_{0}$
is the zeroth Bessel function. Only values of $A/\omega\geq 0$ up to the first minimum of $\mathcal{J}_{0}$ are shown. The vertical axis corresponds to the driving frequency. The phase boundaries are extracted from finite-size
scaling of quasienergy level statistics (see Figs.$\,$2,4, and Supplemental Materials); the solid line is a guide to the eye. Inset: schematic undriven phase diagram
as a function of  disorder $W/J$ and interactions $U/J$, as numerically obtained by \cite{BarLev2015,Bera2015a}. The cross
marks the parameters chosen for our simulations; the red arrow indicates
the effective change of parameters due to the suppression of $J_{eff}$ by the periodic drive.}
\end{figure}

Generically, subjecting an MBL system to a periodic drive increases
the localization length \cite{Ponte2015,Lazarides2015,Abanin2016}.
If the driving is done at sufficiently low frequencies or high amplitudes,
it may even cause the system to exit the MBL phase. This delocalization
effect is caused by transitions such as photon-assisted hopping, which
are mediated by the periodic drive. These transitions conserve energy
only modulo $\hbar\omega$, and can therefore lead to new many-body
resonances which destabilize localization.

An oscillating linear potential (henceforth an AC electric field) has a more subtle effect, as it can effectively
suppress the hopping amplitude between adjacent lattice sites. This
effect, called dynamical localization \cite{Dunlap1986} or coherent
destruction of tunneling \cite{Grossmann1991}, has been implemented
in cold atoms \cite{Lignier2007,Eckardt2009,Eckardt2016}, and can be used, for
example, to induce a transition from a superfluid to
a Mott insulator \cite{Eckardt2005,Zenesini2009}.
In non-interacting systems, dynamical localization can be employed
to tune the localization properties and relaxation dynamics of one-dimensional disordered lattices \cite{Holthaus1995,Hone1993,Martinez2006,Drese1997,Das2015}.
In an interacting disordered system, we expect the suppression of
the hopping amplitude to increase the relative strengths of disorder
and interactions, potentially driving a static delocalized system
towards the MBL phase (Fig. 1, inset). However, it is unclear to what
extent the phenomenon of dynamical localization, obtained to lowest
order in inverse frequency expansions, still applies in interacting
systems, where these expansions often diverge in the thermodynamic
limit at any finite driving frequency \cite{DAlessio2013,DAlessio2014,Bukov2015}.

Here, we demonstrate that an AC electric field affects disordered many-body
systems very differently than generic time-periodic drives. In fact,
we show that an ergodic (delocalized) system subjected to an AC electric
field may transition into the MBL phase. Our results are summarized
in Fig. 1, displaying the phase diagram of a one-dimensional many-body
system which would be ergodic in the absence of driving. Our main
finding is a driving induced MBL phase emerging in a range of driving
amplitudes above a critical driving frequency.

\paragraph{Model.}
We consider interacting hardcore bosons hopping on a disordered one-dimensional lattice with periodic boundary conditions at half filling. The particles hop between neighboring sites with a hopping
amplitude $J$; they interact with nearest-neighbor repulsion $U$,
and are subject to random on-site potentials $V_{i}$ drawn uniformly
and independently from the interval $\left[-W,W\right]$. The static
Hamiltonian in the absence of driving therefore takes the form:

\begin{equation}
\begin{aligned}H_{stat}\left(J,U,W\right)= & J\sum_{i}\left(\hat{c}_{i+1}^{\dagger}\hat{c}_{i}+h.c.\right)+\\
 & +\sum_{i}V_{i}\hat{n}_{i}+U\sum_{i}\hat{n}_{i}\hat{n}_{i+1}
\end{aligned}
\label{eq: H0}
\end{equation}

Variants of this model have been studied extensively \cite{Oganesyan2007,Pal2010,BarLev2015,Bera2015a},
and feature a transition to a many-body localized phase for sufficiently
strong disorder. Specifically, starting from any point in the phase
diagram in the space of normalized disorder $W/J$ and interactions $U/J$, and
decreasing the hopping amplitude $J$, leads to the MBL
phase (Fig. 1).

We investigate the effect of subjecting this static
system to an AC electric field. We work in a gauge where the AC electric field is induced by
a temporally oscillating, spatially uniform vector potential, rather
than by a scalar potential. Using periodic boundary conditions, our
system is thus equivalent to a ring penetrated by an oscillating magnetic
flux. Parameterizing the electric field as $E(t)=A\cos\left(\omega t\right)$,
the Peierls substitution \cite{feynman1966} yields a complex
phase for the hopping amplitude, replacing $\hat{c}_{i+1}^{\dagger}\hat{c}_{i}$
in Eq. (\ref{eq: H0}) with $\hat{c}_{i+1}^{\dagger}\hat{c}_{i}e^{-i(A/\omega)\sin\left(\omega t\right)}$.

Intuitively, the AC field can lead to an effective suppression of the hopping amplitude $J$ due to destructive interference. This effect can be directly seen by considering
the time-averaged Hamiltonian:
\begin{equation}
H_{eff}=\frac{1}{T}\intop_{0}^{T}H_{driven}\left(t\right)dt = H_{stat}\left(J_{eff},U,W\right).
\label{eq: Heff}
\end{equation}
While the disorder and interaction terms remain unchanged,
time-averaging the oscillating phase yields a renormalized effective
hopping amplitude $J_{eff}/J=\mathcal{J}_{0}\left(A/\omega\right)$,
where $\mathcal{J}_{0}$ is the zeroth Bessel function. Expanding in Fourier modes we obtain:
\begin{equation}
H(t)=H_{eff}+\left[\sum_{n\neq0,i}\mathcal{J}_{n}(A/\omega)\hat{c}_{i+1}^{\dagger}\hat{c}_{i}e^{-i\omega nt}+h.c.\right],
\label{eq: hamil_decomposition}
\end{equation}

\noindent where $\mathcal{J}_{n}$ are the Bessel functions of order $n$. For fixed $V$, $U$, we denote by $J_c$ the critical hopping amplitude for localization in the undriven model $H_{stat}$. When $\left|J_{eff}\right|$ is larger
than $J_{c}$, the time-averaged Hamiltonian $H_{eff}$ is delocalized \footnote{$J\mapsto-J$ is a gauge symmetry of Eq. (\ref{eq: H0})}. Therefore, for $\left|J_{eff}\right|>J_c$ we expect the driven system to remain in the delocalized phase.

For $\left|J_{eff}\right|<J_{c}$, the time-averaged Hamiltonian $H_{eff}$ enters
the MBL phase. Since the drive consists of a sum of local and bounded operators ($\sum_{n\neq0}|\mathcal{J}_n|^2\leq1$), energy absorption is suppressed at sufficiently high driving frequencies \cite{Abanin2015}. Therefore, we expect the driven system to become localized above a critical driving frequency \cite{Ponte2015,Abanin2016,Lazarides2015,Das2017}.

To predict the shape of the phase diagram (Fig. 1), we note that eigenstates of
$H_{eff}$ coupled by a local operator can only differ within a range of the order $\xi$ of the operator's
support \cite{Serbyn2013,Huse2014,Ros2014,Chandran2015}, where $\xi$
is the localization length. Absorption of energy from
the drive is therefore expected to be suppressed if the driving frequency $\omega$
is larger than the typical local spectrum of a subsystem of size $\xi$
\cite{Lazarides2015}. Consequently, the critical frequency for
inducing localization should be minimal when the rescaled driving amplitude $A/\omega$ is tuned to a root of the Bessel function ($J_{eff}=0$). In this case, $H_{eff}$ is trivially localized with $\xi=1$, and commutes with the particle occupations $n_{i}$.
With increasing $\left|J_{eff}\right|$, the localization length $\xi$ becomes larger, and the local spectrum grows accordingly \footnote{Tuning $A/\omega$ away from a root of $\mathcal{J}_0$ also decreases the time-dependent terms since $\sum_{n\neq0}|\mathcal{J}_n|^2=1-|\mathcal{J}_0|^2$. However, since the localization length of $H_{eff}$ diverges as $J_{eff}$ approaches $J_c$, we expect it to be the dominant factor in determining the shape of the phase diagram.}. We therefore expect the critical frequency
to increase with $\left|J_{eff}\right|$, until it diverges at $\left|J_{eff}\right|=J_{c}$
where $H_{eff}$ delocalizes.

\paragraph*{Numerical simulations.}
\begin{figure}
\includegraphics[width=8.6cm]{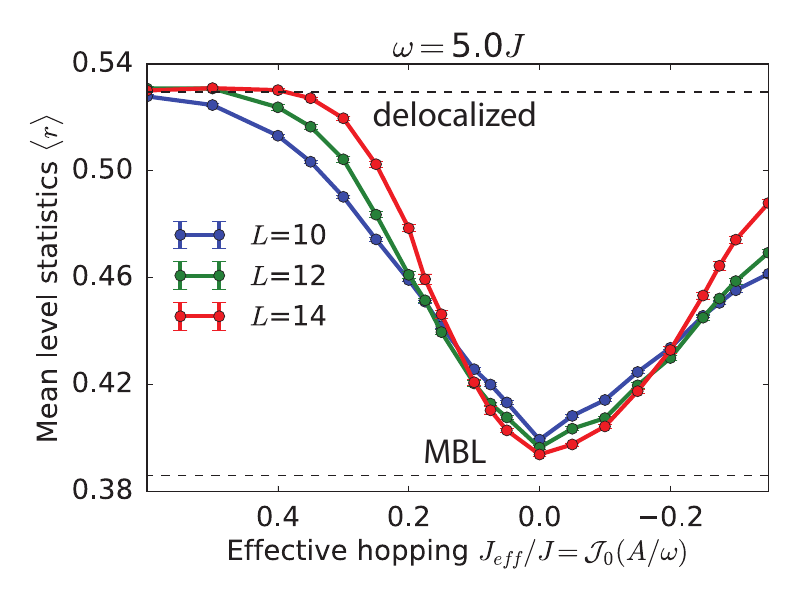}
\protect\caption{Finite-size scaling of quasienergy level statistics as a function
of driving amplitude at high driving frequency ($\omega=5J$). At weak
driving the average level statistics parameter $\left\langle r\right\rangle $
increases with system size and approaches the Wigner-Dyson value $\left\langle r\right\rangle \approx0.53$
corresponding to the delocalized phase. As the driving amplitude increases
and the effective hopping is more strongly suppressed, this trend
is reversed: the level statistics parameter decreases with system
size, approaching the Poisson value $\left\langle r\right\rangle \approx0.39$ which corresponds to the localized phase. Further
increase of driving amplitude leads to a revival of the effective
hopping with an opposite sign, restoring the delocalized phase (only values of $A/\omega\geq 0$ up to the first minimum of $\mathcal{J}_{0}$ are shown). Error
bars indicate one standard deviation of the average, with averaging
performed over at least 1000, 200, 100 disorder realizations for $L=10,12,14$,
respectively.}

\end{figure}

To establish the existence of the driving induced MBL phase, we tune the parameters of our static Hamiltonian to the delocalized
phase: $U=1.5J$, $W=2J$. We then test whether it becomes localized
for various driving frequencies and amplitudes near the first root of the zeroth Bessel function, $\left(A/\omega\right)^*$. Specifically, we examine
the quasienergy level statistics of the evolution operator over one driving period $U\left(T\right)=\mathcal{P}e^{-i\int_{0}^{T}H(t)dt}$
(up to a system size $L=16$), and the relaxation in time of an initially
prepared product state (up to $L=20$). We first establish localization
for strong driving at high frequencies, and then look at the effect
of lower frequencies.

\paragraph*{Finite-size scaling of quasienergy level statistics.} We
compute $U\left(T\right)$
by exponentiating $H\left(t\right)$ at discrete time-steps (120 equally
spaced steps) using exact diagonalization (ED). We then diagonalize
$U\left(T\right)$ to obtain the quasienergies $\epsilon_{\alpha}$,
and focus on the gaps between subsequent quasienergies $\delta_{\alpha}=\epsilon_{\alpha+1}-\epsilon_{\alpha}$.
The ratio between subsequent gaps averaged over the quasienergy spectrum
$\left\langle r\right\rangle =\left\langle\frac{\min\left(\delta_{\alpha},\delta_{\alpha+1}\right)}{\max\left(\delta_{\alpha},\delta_{\alpha+1}\right)}\right\rangle$
measures the repulsion between quasienergy levels, and distinguishes between the
MBL and ergodic phases of a driven system. The ergodic, delocalized phase exhibits quasienergy level repulsion with a level spacings parameter $\left\langle r\right\rangle _{COE}\approx0.53$
corresponding to the circular orthogonal ensemble (COE) of random
matrices \cite{DAlessio2014}. The MBL phase features uncorrelated Poisson quasienergy level
statistics and therefore a smaller value $\left\langle r\right\rangle _{POI}\approx0.39$ \cite{Oganesyan2007,Ponte2015,Zhang2016}.

We begin by considering a fixed frequency $\omega=5J$ and increasing
driving amplitudes (see Fig 2). We observe a sharp change in the scaling
of the level statistics with the system size as we increase the driving
amplitude. When the system is weakly driven, the level statistics
parameter becomes larger as the system size is increased, approaching
the delocalized value $\left\langle r\right\rangle _{COE}\approx0.53$
in a similar manner to the undriven case \cite{BarLev2015}. However, at sufficiently
strong driving the effective hopping $J_{eff}/J=\mathcal{J}_{0}\left(A/\omega\right)$
is suppressed, and this trend is reversed: the level statistics parameter
 decreases as the system size is increased, approaching the
MBL value $\left\langle r\right\rangle _{POI}\approx0.39$. Thus,
the drive induces a transition from the delocalized phase into the
MBL phase. At even stronger driving amplitudes $\left|J_{eff}\right|$ rises
again, and the delocalized phase is recovered.
We estimate the critical values of $J_{eff}$ for the transitions between the MBL and ergodic phases (marked
in Fig. 1) at the crossings of the
curves for different system sizes $L$ according to finite-size scaling (see Supplemental Materials).

The width (standard deviation) of the many-body spectrum of $H_{stat}$ in the system sizes we studied with ED is comparable to the driving frequency. This renders resonant absorption of energy from the drive less prominent than in the thermodynamic limit. To confirm the existence of the driving induced localized phase, we study larger systems by propagating an initial density pattern in time.

\begin{figure*}
\includegraphics[width=18cm]{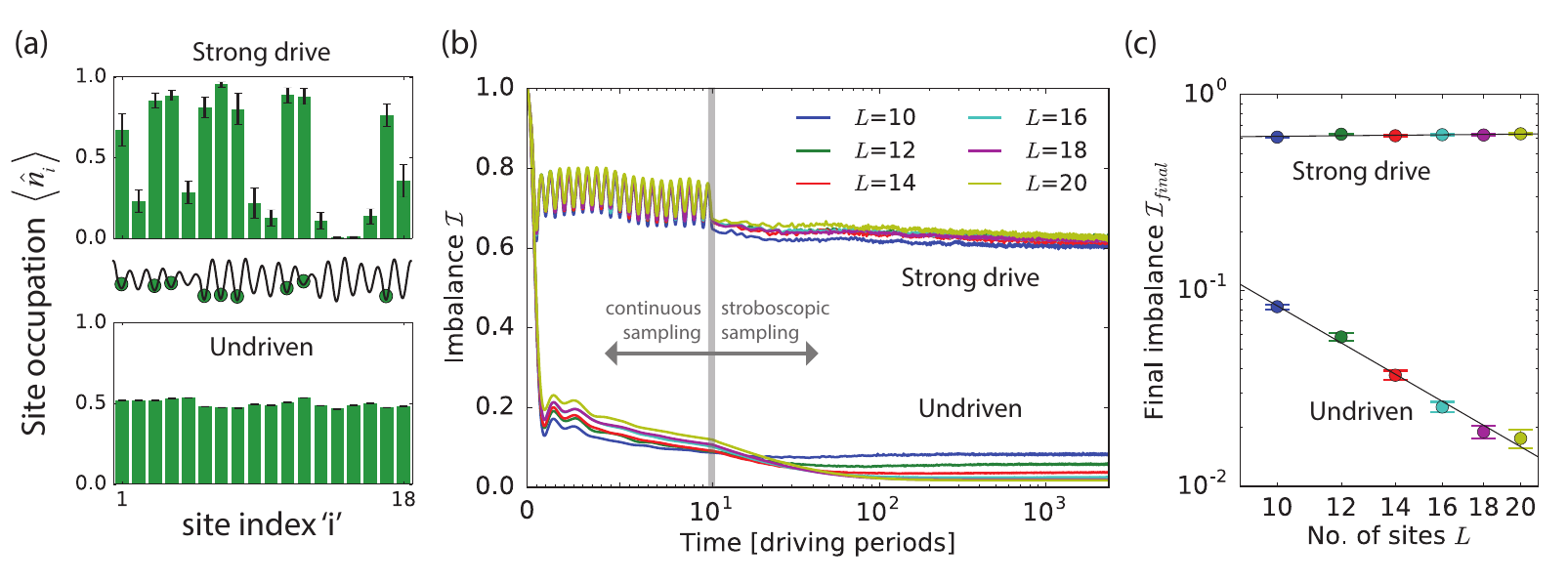}

\protect\caption{Relaxation of the occupation imbalance starting from a product state.
(a) Example of site occupations $\left\langle n_{i}\right\rangle $
at long times with a strong high-frequency drive (top; $\omega=5J$,
$A/\omega=(A/\omega)^*$ corresponding to $J_{eff}=0$) vs. an undriven system (bottom) starting from the same
initial product state and disorder realization (middle). The occupations are averaged over
$10^{3}T<t<1.5\times10^{3}T$ with error bars indicating one standard
deviation accounting for fluctuations in that duration. (b) Imbalance
[see Eq. (\ref{eq: imbalance})] as a function of time with (top)
vs. without (bottom) a strong high-frequency drive ($\omega=5J$, $J_{eff}=0$), averaged over disorder realizations and random initial product states ($1000$ instances for $L=10$, $500$ for $12\leq L\leq18$ and $200$
for $L=20$). The imbalance is measured at each simulation step and
shown on a linear time scale up to $t=10T$ , after which it is measured
at stroboscopic times $t=nT$ only and shown on a logarithmic time
scale. While in the absence of driving site occupations become uncorrelated
with their initial values, in the driven case they remain highly correlated.
(c) Final imbalance as a function of system size (log-log scale),
averaged over $1.4\times10^{3}T<t<1.5\times10^{3}T$; error bars indicate one
standard deviation of the average over disorder realizations. The final imbalance decreases with
system size in the absence of driving (slope$=-2.4\pm0.2$), whereas
it is insensitive to system size when the drive is applied (slope$=0.04\pm0.03$).}
\end{figure*}

\begin{figure}[b]
\includegraphics[width=8.6cm]{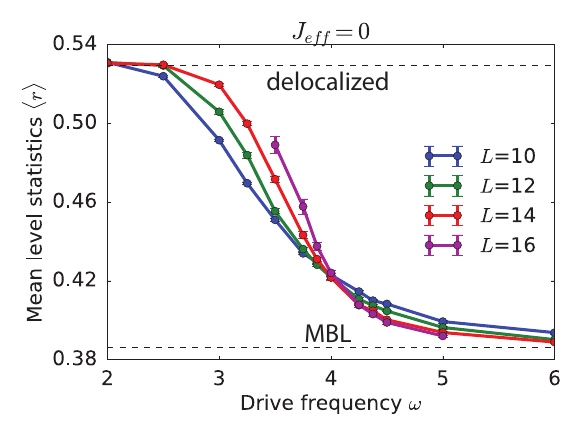}
\protect\caption{Quasienergy level statistics $\left\langle r\right\rangle$
as a function of driving frequency for $A/\omega=(A/\omega)^*$ corresponding to $J_{eff}=0$. The
changing trend in the scaling of $\left\langle r\right\rangle $ with
system size indicates a critical frequency $\omega_{c}$ ($\approx4J$
for our parameters) above which driving induced localization occurs.}
\end{figure}

\paragraph{Relaxation of an initial product state. }We initialize our
system in an arbitrary product state of site occupations, distributing
$L/2$ particles randomly among its $L$ sites. We then evolve this state for a long time by acting
on it with the exponential of the Hamiltonian at discrete time steps
(120 steps per period for over 1500 driving periods), and follow the
site occupations $\left\langle \hat{n}_{i}\left(t\right)\right\rangle $.
In the absence of driving, the particles spread throughout the system,
such that the occupation in each site eventually revolves around $\left\langle \hat{n}_{i}\right\rangle \approx0.5$
(Fig. 3a, bottom), as expected for an ergodic system \cite{SrednickiETH,Iyer2013}.

When we evolve the same state
with a strong drive at high frequency ($\omega=5J$), the particles
remain mainly in their initial positions for the duration of our simulations (Fig. 3a, top).
This indicates long-term memory of the initial conditions, a signature
of the MBL phase, as expected from the ED results. Following \cite{Edwards1975,Iyer2013,Schreiber2015,Gopalakrishnan2016a,Bordia2017,Luitz2016a},
we differentiate between the MBL and delocalized phases by tracking the evolution
of the generalized imbalance,
which measures the correlation between the current and initial density
patterns,

\begin{equation}
\mathcal{I}\left(t\right)=\frac{4}{L}\sum_{i=1}^{L}\broket{\psi\left(0\right)}{\left(\hat{n}_{i}\left(t\right)-\frac{1}{2}\right)\left(\hat{n}_{i}\left(0\right)-\frac{1}{2}\right)}{\psi\left(0\right)}.
\label{eq: imbalance}
\end{equation}
The imbalance generalizes a technique used in recent cold atom experiments,
which studied the relaxation of an initially prepared charge-density-wave
\cite{Schreiber2015,Bordia2017}.

In the absence of driving, the system is ergodic, and the density
pattern becomes uncorrelated with the initial pattern. Thus, the imbalance
decays to a value which decreases with system size (Fig. 3b, c). When
a drive of appropriate frequency and amplitude is applied, memory of the initial occupancy pattern persists
for long times, and the imbalance stabilizes on a finite value independent
of the system size ($\mathcal{I}\approx0.6$ for $\omega=5J$, $J_{eff}=0$).
Thus, in the strong driving regime the system fails to thermalize,
indicating that an MBL phase is induced by the driving field.

\paragraph*{Critical driving frequency.} Tuning the rescaled driving amplitude $A/\omega$ to
the first root of $\mathcal{J}_0$ (such that $J_{eff}=0$) and repeating the level
statistics analysis for varying frequencies, we find a minimal critical frequency $\omega_c\approx4J$
for inducing localization with the driving field and model parameters we considered. Below this frequency, the level statistics parameter $\left\langle r\right\rangle $
tends to its delocalized value as the system size increases (Fig. 4). Ergodicity at $\omega=3.5J$ is further confirmed by the analysis of the long-time imbalance of initial product states (Fig. S1 in Supplemental Materials).

As discussed above, we expect the critical frequency for inducing the MBL phase in our system to increase with $\left|J_{eff}\right|$. This expectation is indeed confirmed by the phase diagram in Fig. 1, which we obtained by analyzing the level statistics for additional cuts of fixed frequency in parameter space (see Fig. S2 for data near $\omega_c$). Finally, we find qualitatively similar results with a slightly reduced critical frequency when the driving amplitude is tuned near the second root of $\mathcal{J}_0$; when the system is taken at a smaller filling fraction; or when a square-wave drive is used (for details, see Supplemental Materials).

\paragraph*{Discussion.}
We have shown that subjecting an ergodic system to a periodic drive
can induce a transition into the MBL phase, providing an interesting
example for the emergence of integrability in an ergodic system due
to the addition of a drive. It would be interesting to understand
how the phase diagram (Fig. 1) depends on the strength of disorder
and interactions, and specifically, whether MBL can be induced
in our model starting from arbitrarily weak disorder. Especially interesting
are possible generalizations to higher dimensions, for example by using
circularly polarized electromagnetic fields in two dimensions \cite{Bukov2015a}.
Most importantly, our results open new possibilities for inducing
exotic out-of equilibrium phases in weakly disordered systems using
methods which are readily accessible in cold atom systems \cite{Lignier2007,Eckardt2016,Bordia2017}.

\textit{Note added:} During the completion of this manuscript, we became aware of recent works which find localization enhancement in the driven quantum random energy model \cite{Burin2017}, and a driving induced MBL phase in spin chains \cite{Soonwon2017}.

\begin{acknowledgments}
We thank Dima Abanin, Jens Bardarson, Iliya Esin, Vladimir Kalnizky, Ilia Khait, Achilleas Lazarides, Roderich Moessner and Alon Nahshony for illuminating
discussions. E. B. acknowledges financial support from the Gutwirth Foundation. G. R. is grateful for support from the NSF through Grant
No. DMR-1410435, the Institute of Quantum Information and Matter, an NSF Frontier center funded by the
Gordon and Betty Moore Foundation, and the Packard
Foundation. N. L. acknowledges support from the People Programme
(Marie Curie Actions) of the European Union\textquoteright s Seventh
Framework Programme (No. FP7/2007\textendash 2013) under REA Grant
Agreement No. 631696, from the Israeli Center of Research Excellence
(I-CORE) \textquotedblleft Circle of Light.\textquotedblright{},
and from the European Research Council (ERC) under the European Union
Horizon 2020 Research and Innovation Programme (Grant Agreement No.
639172).
\end{acknowledgments}

\bibliographystyle{apsrev4-1}

\begin{thebibliography}{67}%
\makeatletter
\providecommand \@ifxundefined [1]{%
 \@ifx{#1\undefined}
}%
\providecommand \@ifnum [1]{%
 \ifnum #1\expandafter \@firstoftwo
 \else \expandafter \@secondoftwo
 \fi
}%
\providecommand \@ifx [1]{%
 \ifx #1\expandafter \@firstoftwo
 \else \expandafter \@secondoftwo
 \fi
}%
\providecommand \natexlab [1]{#1}%
\providecommand \enquote  [1]{``#1''}%
\providecommand \bibnamefont  [1]{#1}%
\providecommand \bibfnamefont [1]{#1}%
\providecommand \citenamefont [1]{#1}%
\providecommand \href@noop [0]{\@secondoftwo}%
\providecommand \href [0]{\begingroup \@sanitize@url \@href}%
\providecommand \@href[1]{\@@startlink{#1}\@@href}%
\providecommand \@@href[1]{\endgroup#1\@@endlink}%
\providecommand \@sanitize@url [0]{\catcode `\\12\catcode `\$12\catcode
  `\&12\catcode `\#12\catcode `\^12\catcode `\_12\catcode `\%12\relax}%
\providecommand \@@startlink[1]{}%
\providecommand \@@endlink[0]{}%
\providecommand \url  [0]{\begingroup\@sanitize@url \@url }%
\providecommand \@url [1]{\endgroup\@href {#1}{\urlprefix }}%
\providecommand \urlprefix  [0]{URL }%
\providecommand \Eprint [0]{\href }%
\providecommand \doibase [0]{http://dx.doi.org/}%
\providecommand \selectlanguage [0]{\@gobble}%
\providecommand \bibinfo  [0]{\@secondoftwo}%
\providecommand \bibfield  [0]{\@secondoftwo}%
\providecommand \translation [1]{[#1]}%
\providecommand \BibitemOpen [0]{}%
\providecommand \bibitemStop [0]{}%
\providecommand \bibitemNoStop [0]{.\EOS\space}%
\providecommand \EOS [0]{\spacefactor3000\relax}%
\providecommand \BibitemShut  [1]{\csname bibitem#1\endcsname}%
\let\auto@bib@innerbib\@empty
\bibitem [{\citenamefont {D'Alessio}\ and\ \citenamefont
  {Rigol}(2014)}]{DAlessio2014}%
  \BibitemOpen
  \bibfield  {author} {\bibinfo {author} {\bibfnamefont {L.}~\bibnamefont
  {D'Alessio}}\ and\ \bibinfo {author} {\bibfnamefont {M.}~\bibnamefont
  {Rigol}},\ }\href@noop {} {\bibfield  {journal} {\bibinfo  {journal} {Phys.
  Rev. X}\ }\textbf {\bibinfo {volume} {4}},\ \bibinfo {pages} {041048}
  (\bibinfo {year} {2014})}\BibitemShut {NoStop}%
\bibitem [{\citenamefont {Lazarides}\ \emph
  {et~al.}(2014{\natexlab{a}})\citenamefont {Lazarides}, \citenamefont {Das},\
  and\ \citenamefont {Moessner}}]{Lazarides2014}%
  \BibitemOpen
  \bibfield  {author} {\bibinfo {author} {\bibfnamefont {A.}~\bibnamefont
  {Lazarides}}, \bibinfo {author} {\bibfnamefont {A.}~\bibnamefont {Das}}, \
  and\ \bibinfo {author} {\bibfnamefont {R.}~\bibnamefont {Moessner}},\
  }\href@noop {} {\bibfield  {journal} {\bibinfo  {journal} {Phys. Rev. E}\
  }\textbf {\bibinfo {volume} {90}},\ \bibinfo {pages} {012110} (\bibinfo
  {year} {2014}{\natexlab{a}})}\BibitemShut {NoStop}%
\bibitem [{\citenamefont {Lazarides}\ \emph
  {et~al.}(2014{\natexlab{b}})\citenamefont {Lazarides}, \citenamefont {Das},\
  and\ \citenamefont {Moessner}}]{Lazarides2014a}%
  \BibitemOpen
  \bibfield  {author} {\bibinfo {author} {\bibfnamefont {A.}~\bibnamefont
  {Lazarides}}, \bibinfo {author} {\bibfnamefont {A.}~\bibnamefont {Das}}, \
  and\ \bibinfo {author} {\bibfnamefont {R.}~\bibnamefont {Moessner}},\ }\href
  {\doibase 10.1103/PhysRevLett.112.150401} {\bibfield  {journal} {\bibinfo
  {journal} {Phys. Rev. Lett.}\ }\textbf {\bibinfo {volume} {112}},\ \bibinfo
  {pages} {150401} (\bibinfo {year} {2014}{\natexlab{b}})}\BibitemShut
  {NoStop}%
\bibitem [{\citenamefont {Chandran}\ and\ \citenamefont
  {Sondhi}(2016)}]{Chandran2016}%
  \BibitemOpen
  \bibfield  {author} {\bibinfo {author} {\bibfnamefont {A.}~\bibnamefont
  {Chandran}}\ and\ \bibinfo {author} {\bibfnamefont {S.~L.}\ \bibnamefont
  {Sondhi}},\ }\href@noop {} {\bibfield  {journal} {\bibinfo  {journal} {Phys.
  Rev. B}\ }\textbf {\bibinfo {volume} {93}},\ \bibinfo {pages} {174305}
  (\bibinfo {year} {2016})}\BibitemShut {NoStop}%
\bibitem [{\citenamefont {Citro}\ \emph {et~al.}(2015)\citenamefont {Citro},
  \citenamefont {{Dalla Torre}}, \citenamefont {D'Alessio}, \citenamefont
  {Polkovnikov}, \citenamefont {Babadi}, \citenamefont {Oka},\ and\
  \citenamefont {Demler}}]{Citro2015}%
  \BibitemOpen
  \bibfield  {author} {\bibinfo {author} {\bibfnamefont {R.}~\bibnamefont
  {Citro}}, \bibinfo {author} {\bibfnamefont {E.~G.}\ \bibnamefont {{Dalla
  Torre}}}, \bibinfo {author} {\bibfnamefont {L.}~\bibnamefont {D'Alessio}},
  \bibinfo {author} {\bibfnamefont {A.}~\bibnamefont {Polkovnikov}}, \bibinfo
  {author} {\bibfnamefont {M.}~\bibnamefont {Babadi}}, \bibinfo {author}
  {\bibfnamefont {T.}~\bibnamefont {Oka}}, \ and\ \bibinfo {author}
  {\bibfnamefont {E.}~\bibnamefont {Demler}},\ }\href@noop {} {\bibfield
  {journal} {\bibinfo  {journal} {Annals of Physics}\ }\textbf {\bibinfo
  {volume} {360}},\ \bibinfo {pages} {694} (\bibinfo {year}
  {2015})}\BibitemShut {NoStop}%
\bibitem [{\citenamefont {Kukuljan}\ and\ \citenamefont
  {Prosen}(2016)}]{Kukuljan2016}%
  \BibitemOpen
  \bibfield  {author} {\bibinfo {author} {\bibfnamefont {I.}~\bibnamefont
  {Kukuljan}}\ and\ \bibinfo {author} {\bibfnamefont {T.}~\bibnamefont
  {Prosen}},\ }\href@noop {} {\bibfield  {journal} {\bibinfo  {journal}
  {Journal of Statistical Mechanics: Theory and Experiment}\ }\textbf {\bibinfo
  {volume} {2016}},\ \bibinfo {pages} {043305} (\bibinfo {year}
  {2016})}\BibitemShut {NoStop}%
\bibitem [{\citenamefont {D'Alessio}\ and\ \citenamefont
  {Polkovnikov}(2013)}]{DAlessio2013}%
  \BibitemOpen
  \bibfield  {author} {\bibinfo {author} {\bibfnamefont {L.}~\bibnamefont
  {D'Alessio}}\ and\ \bibinfo {author} {\bibfnamefont {A.}~\bibnamefont
  {Polkovnikov}},\ }\href@noop {} {\bibfield  {journal} {\bibinfo  {journal}
  {Annals of Physics}\ }\textbf {\bibinfo {volume} {333}},\ \bibinfo {pages}
  {19} (\bibinfo {year} {2013})}\BibitemShut {NoStop}%
\bibitem [{\citenamefont {Ponte}\ \emph
  {et~al.}(2015{\natexlab{a}})\citenamefont {Ponte}, \citenamefont {Chandran},
  \citenamefont {Papi{\'{c}}},\ and\ \citenamefont {Abanin}}]{Ponte2015a}%
  \BibitemOpen
  \bibfield  {author} {\bibinfo {author} {\bibfnamefont {P.}~\bibnamefont
  {Ponte}}, \bibinfo {author} {\bibfnamefont {A.}~\bibnamefont {Chandran}},
  \bibinfo {author} {\bibfnamefont {Z.}~\bibnamefont {Papi{\'{c}}}}, \ and\
  \bibinfo {author} {\bibfnamefont {D.~A.}\ \bibnamefont {Abanin}},\
  }\href@noop {} {\bibfield  {journal} {\bibinfo  {journal} {Annals of
  Physics}\ }\textbf {\bibinfo {volume} {353}},\ \bibinfo {pages} {196}
  (\bibinfo {year} {2015}{\natexlab{a}})}\BibitemShut {NoStop}%
\bibitem [{\citenamefont {Lazarides}\ \emph {et~al.}(2015)\citenamefont
  {Lazarides}, \citenamefont {Das},\ and\ \citenamefont
  {Moessner}}]{Lazarides2015}%
  \BibitemOpen
  \bibfield  {author} {\bibinfo {author} {\bibfnamefont {A.}~\bibnamefont
  {Lazarides}}, \bibinfo {author} {\bibfnamefont {A.}~\bibnamefont {Das}}, \
  and\ \bibinfo {author} {\bibfnamefont {R.}~\bibnamefont {Moessner}},\
  }\href@noop {} {\bibfield  {journal} {\bibinfo  {journal} {Phys. Rev. Lett.}\
  }\textbf {\bibinfo {volume} {115}},\ \bibinfo {pages} {030402} (\bibinfo
  {year} {2015})}\BibitemShut {NoStop}%
\bibitem [{\citenamefont {Ponte}\ \emph
  {et~al.}(2015{\natexlab{b}})\citenamefont {Ponte}, \citenamefont
  {Papi{\'{c}}}, \citenamefont {Huveneers},\ and\ \citenamefont
  {Abanin}}]{Ponte2015}%
  \BibitemOpen
  \bibfield  {author} {\bibinfo {author} {\bibfnamefont {P.}~\bibnamefont
  {Ponte}}, \bibinfo {author} {\bibfnamefont {Z.}~\bibnamefont {Papi{\'{c}}}},
  \bibinfo {author} {\bibfnamefont {F.}~\bibnamefont {Huveneers}}, \ and\
  \bibinfo {author} {\bibfnamefont {D.~A.}\ \bibnamefont {Abanin}},\
  }\href@noop {} {\bibfield  {journal} {\bibinfo  {journal} {Phys. Rev. Lett.}\
  }\textbf {\bibinfo {volume} {114}},\ \bibinfo {pages} {140401} (\bibinfo
  {year} {2015}{\natexlab{b}})}\BibitemShut {NoStop}%
\bibitem [{\citenamefont {Abanin}\ \emph {et~al.}(2016)\citenamefont {Abanin},
  \citenamefont {{De Roeck}},\ and\ \citenamefont {Huveneers}}]{Abanin2016}%
  \BibitemOpen
  \bibfield  {author} {\bibinfo {author} {\bibfnamefont {D.~A.}\ \bibnamefont
  {Abanin}}, \bibinfo {author} {\bibfnamefont {W.}~\bibnamefont {{De Roeck}}},
  \ and\ \bibinfo {author} {\bibfnamefont {F.}~\bibnamefont {Huveneers}},\
  }\href@noop {} {\bibfield  {journal} {\bibinfo  {journal} {Annals of
  Physics}\ }\textbf {\bibinfo {volume} {372}},\ \bibinfo {pages} {1} (\bibinfo
  {year} {2016})}\BibitemShut {NoStop}%
\bibitem [{\citenamefont {Rehn}\ \emph {et~al.}(2016)\citenamefont {Rehn},
  \citenamefont {Lazarides}, \citenamefont {Pollmann},\ and\ \citenamefont
  {Moessner}}]{Rehn2016}%
  \BibitemOpen
  \bibfield  {author} {\bibinfo {author} {\bibfnamefont {J.}~\bibnamefont
  {Rehn}}, \bibinfo {author} {\bibfnamefont {A.}~\bibnamefont {Lazarides}},
  \bibinfo {author} {\bibfnamefont {F.}~\bibnamefont {Pollmann}}, \ and\
  \bibinfo {author} {\bibfnamefont {R.}~\bibnamefont {Moessner}},\ }\href@noop
  {} {\bibfield  {journal} {\bibinfo  {journal} {Phys. Rev. B}\ }\textbf
  {\bibinfo {volume} {94}},\ \bibinfo {pages} {020201} (\bibinfo {year}
  {2016})}\BibitemShut {NoStop}%
\bibitem [{\citenamefont {Gopalakrishnan}\ \emph {et~al.}(2016)\citenamefont
  {Gopalakrishnan}, \citenamefont {Knap},\ and\ \citenamefont
  {Demler}}]{Gopalakrishnan2016a}%
  \BibitemOpen
  \bibfield  {author} {\bibinfo {author} {\bibfnamefont {S.}~\bibnamefont
  {Gopalakrishnan}}, \bibinfo {author} {\bibfnamefont {M.}~\bibnamefont
  {Knap}}, \ and\ \bibinfo {author} {\bibfnamefont {E.}~\bibnamefont
  {Demler}},\ }\href@noop {} {\bibfield  {journal} {\bibinfo  {journal} {Phys.
  Rev. B}\ }\textbf {\bibinfo {volume} {94}},\ \bibinfo {pages} {094201}
  (\bibinfo {year} {2016})}\BibitemShut {NoStop}%
\bibitem [{\citenamefont {Anderson}(1958)}]{Anderson1958}%
  \BibitemOpen
  \bibfield  {author} {\bibinfo {author} {\bibfnamefont {P.~W.}\ \bibnamefont
  {Anderson}},\ }\href@noop {} {\bibfield  {journal} {\bibinfo  {journal}
  {Phys. Rev.}\ }\textbf {\bibinfo {volume} {109}},\ \bibinfo {pages} {1492}
  (\bibinfo {year} {1958})}\BibitemShut {NoStop}%
\bibitem [{\citenamefont {Basko}\ \emph {et~al.}(2006)\citenamefont {Basko},
  \citenamefont {Aleiner},\ and\ \citenamefont {Altshuler}}]{Basko2006}%
  \BibitemOpen
  \bibfield  {author} {\bibinfo {author} {\bibfnamefont {D.~M.}\ \bibnamefont
  {Basko}}, \bibinfo {author} {\bibfnamefont {I.~L.}\ \bibnamefont {Aleiner}},
  \ and\ \bibinfo {author} {\bibfnamefont {B.~L.}\ \bibnamefont {Altshuler}},\
  }\href@noop {} {\bibfield  {journal} {\bibinfo  {journal} {Annals of
  Physics}\ }\textbf {\bibinfo {volume} {321}},\ \bibinfo {pages} {1126}
  (\bibinfo {year} {2006})}\BibitemShut {NoStop}%
\bibitem [{\citenamefont {Gornyi}\ \emph {et~al.}(2005)\citenamefont {Gornyi},
  \citenamefont {Mirlin},\ and\ \citenamefont {Polyakov}}]{Gornyi2005}%
  \BibitemOpen
  \bibfield  {author} {\bibinfo {author} {\bibfnamefont {I.~V.}\ \bibnamefont
  {Gornyi}}, \bibinfo {author} {\bibfnamefont {A.~D.}\ \bibnamefont {Mirlin}},
  \ and\ \bibinfo {author} {\bibfnamefont {D.~G.}\ \bibnamefont {Polyakov}},\
  }\href@noop {} {\bibfield  {journal} {\bibinfo  {journal} {Phys. Rev. Lett.}\
  }\textbf {\bibinfo {volume} {95}},\ \bibinfo {pages} {206603} (\bibinfo
  {year} {2005})}\BibitemShut {NoStop}%
\bibitem [{\citenamefont {Oganesyan}\ and\ \citenamefont
  {Huse}(2007)}]{Oganesyan2007}%
  \BibitemOpen
  \bibfield  {author} {\bibinfo {author} {\bibfnamefont {V.}~\bibnamefont
  {Oganesyan}}\ and\ \bibinfo {author} {\bibfnamefont {D.~A.}\ \bibnamefont
  {Huse}},\ }\href@noop {} {\bibfield  {journal} {\bibinfo  {journal} {Phys.
  Rev. B}\ }\textbf {\bibinfo {volume} {75}},\ \bibinfo {pages} {155111}
  (\bibinfo {year} {2007})}\BibitemShut {NoStop}%
\bibitem [{\citenamefont {Nandkishore}\ and\ \citenamefont
  {Huse}(2015)}]{Nandkishore2015}%
  \BibitemOpen
  \bibfield  {author} {\bibinfo {author} {\bibfnamefont {R.}~\bibnamefont
  {Nandkishore}}\ and\ \bibinfo {author} {\bibfnamefont {D.~A.}\ \bibnamefont
  {Huse}},\ }\href@noop {} {\bibfield  {journal} {\bibinfo  {journal} {Annual
  Review of Condensed Matter Physics}\ }\textbf {\bibinfo {volume} {6}},\
  \bibinfo {pages} {15} (\bibinfo {year} {2015})}\BibitemShut {NoStop}%
\bibitem [{\citenamefont {Moessner}\ and\ \citenamefont
  {Sondhi}()}]{Moessner2017}%
  \BibitemOpen
  \bibfield  {author} {\bibinfo {author} {\bibfnamefont {R.}~\bibnamefont
  {Moessner}}\ and\ \bibinfo {author} {\bibfnamefont {S.~L.}\ \bibnamefont
  {Sondhi}},\ }\href@noop {} {\ }\Eprint {http://arxiv.org/abs/1701.08056}
  {arXiv:1701.08056} \BibitemShut {NoStop}%
\bibitem [{\citenamefont {Titum}\ \emph {et~al.}(2016)\citenamefont {Titum},
  \citenamefont {Berg}, \citenamefont {Rudner}, \citenamefont {Refael},\ and\
  \citenamefont {Lindner}}]{Titum2016}%
  \BibitemOpen
  \bibfield  {author} {\bibinfo {author} {\bibfnamefont {P.}~\bibnamefont
  {Titum}}, \bibinfo {author} {\bibfnamefont {E.}~\bibnamefont {Berg}},
  \bibinfo {author} {\bibfnamefont {M.~S.}\ \bibnamefont {Rudner}}, \bibinfo
  {author} {\bibfnamefont {G.}~\bibnamefont {Refael}}, \ and\ \bibinfo {author}
  {\bibfnamefont {N.~H.}\ \bibnamefont {Lindner}},\ }\href@noop {} {\bibfield
  {journal} {\bibinfo  {journal} {Phys. Rev. X}\ }\textbf {\bibinfo {volume}
  {6}},\ \bibinfo {pages} {021013} (\bibinfo {year} {2016})}\BibitemShut
  {NoStop}%
\bibitem [{\citenamefont {Khemani}\ \emph {et~al.}(2016)\citenamefont
  {Khemani}, \citenamefont {Lazarides}, \citenamefont {Moessner},\ and\
  \citenamefont {Sondhi}}]{PhysRevLett.116.250401}%
  \BibitemOpen
  \bibfield  {author} {\bibinfo {author} {\bibfnamefont {V.}~\bibnamefont
  {Khemani}}, \bibinfo {author} {\bibfnamefont {A.}~\bibnamefont {Lazarides}},
  \bibinfo {author} {\bibfnamefont {R.}~\bibnamefont {Moessner}}, \ and\
  \bibinfo {author} {\bibfnamefont {S.~L.}\ \bibnamefont {Sondhi}},\
  }\href@noop {} {\bibfield  {journal} {\bibinfo  {journal} {Phys. Rev. Lett.}\
  }\textbf {\bibinfo {volume} {116}},\ \bibinfo {pages} {250401} (\bibinfo
  {year} {2016})}\BibitemShut {NoStop}%
\bibitem [{\citenamefont {{Von Keyserlingk}}\ \emph {et~al.}(2016)\citenamefont
  {{Von Keyserlingk}}, \citenamefont {Khemani},\ and\ \citenamefont
  {Sondhi}}]{VonKeyserlingk2016a}%
  \BibitemOpen
  \bibfield  {author} {\bibinfo {author} {\bibfnamefont {C.~W.}\ \bibnamefont
  {{Von Keyserlingk}}}, \bibinfo {author} {\bibfnamefont {V.}~\bibnamefont
  {Khemani}}, \ and\ \bibinfo {author} {\bibfnamefont {S.~L.}\ \bibnamefont
  {Sondhi}},\ }\href@noop {} {\bibfield  {journal} {\bibinfo  {journal} {Phys.
  Rev. B}\ }\textbf {\bibinfo {volume} {94}},\ \bibinfo {pages} {085112}
  (\bibinfo {year} {2016})}\BibitemShut {NoStop}%
\bibitem [{\citenamefont {{Von Keyserlingk}}\ and\ \citenamefont
  {Sondhi}(2016)}]{VonKeyserlingk2016b}%
  \BibitemOpen
  \bibfield  {author} {\bibinfo {author} {\bibfnamefont {C.~W.}\ \bibnamefont
  {{Von Keyserlingk}}}\ and\ \bibinfo {author} {\bibfnamefont {S.~L.}\
  \bibnamefont {Sondhi}},\ }\href@noop {} {\bibfield  {journal} {\bibinfo
  {journal} {Phys. Rev. B}\ }\textbf {\bibinfo {volume} {93}},\ \bibinfo
  {pages} {245145} (\bibinfo {year} {2016})}\BibitemShut {NoStop}%
\bibitem [{\citenamefont {von Keyserlingk}\ and\ \citenamefont
  {Sondhi}(2016)}]{VonKeyserlingk2016}%
  \BibitemOpen
  \bibfield  {author} {\bibinfo {author} {\bibfnamefont {C.~W.}\ \bibnamefont
  {von Keyserlingk}}\ and\ \bibinfo {author} {\bibfnamefont {S.~L.}\
  \bibnamefont {Sondhi}},\ }\href {\doibase 10.1103/PhysRevB.93.245146}
  {\bibfield  {journal} {\bibinfo  {journal} {Phys. Rev. B}\ }\textbf {\bibinfo
  {volume} {93}},\ \bibinfo {pages} {245146} (\bibinfo {year}
  {2016})}\BibitemShut {NoStop}%
\bibitem [{\citenamefont {Else}\ and\ \citenamefont {Nayak}(2016)}]{Else2016}%
  \BibitemOpen
  \bibfield  {author} {\bibinfo {author} {\bibfnamefont {D.~V.}\ \bibnamefont
  {Else}}\ and\ \bibinfo {author} {\bibfnamefont {C.}~\bibnamefont {Nayak}},\
  }\href@noop {} {\bibfield  {journal} {\bibinfo  {journal} {Phys. Rev. B}\
  }\textbf {\bibinfo {volume} {93}},\ \bibinfo {pages} {201103} (\bibinfo
  {year} {2016})}\BibitemShut {NoStop}%
\bibitem [{\citenamefont {Else}\ \emph {et~al.}(2016)\citenamefont {Else},
  \citenamefont {Bauer},\ and\ \citenamefont {Nayak}}]{Else2016a}%
  \BibitemOpen
  \bibfield  {author} {\bibinfo {author} {\bibfnamefont {D.~V.}\ \bibnamefont
  {Else}}, \bibinfo {author} {\bibfnamefont {B.}~\bibnamefont {Bauer}}, \ and\
  \bibinfo {author} {\bibfnamefont {C.}~\bibnamefont {Nayak}},\ }\href@noop {}
  {\bibfield  {journal} {\bibinfo  {journal} {Phys. Rev. Lett.}\ }\textbf
  {\bibinfo {volume} {117}},\ \bibinfo {pages} {090402} (\bibinfo {year}
  {2016})}\BibitemShut {NoStop}%
\bibitem [{\citenamefont {Potirniche}\ \emph {et~al.}()\citenamefont
  {Potirniche}, \citenamefont {Potter}, \citenamefont {Schleier-Smith},
  \citenamefont {Vishwanath},\ and\ \citenamefont {Yao}}]{Potirniche2016}%
  \BibitemOpen
  \bibfield  {author} {\bibinfo {author} {\bibfnamefont {I.-D.}\ \bibnamefont
  {Potirniche}}, \bibinfo {author} {\bibfnamefont {A.~C.}\ \bibnamefont
  {Potter}}, \bibinfo {author} {\bibfnamefont {M.}~\bibnamefont
  {Schleier-Smith}}, \bibinfo {author} {\bibfnamefont {A.}~\bibnamefont
  {Vishwanath}}, \ and\ \bibinfo {author} {\bibfnamefont {N.~Y.}\ \bibnamefont
  {Yao}},\ }\href@noop {} {\ }\Eprint {http://arxiv.org/abs/1610.07611}
  {arXiv:1610.07611} \BibitemShut {NoStop}%
\bibitem [{\citenamefont {Potter}\ \emph {et~al.}(2016)\citenamefont {Potter},
  \citenamefont {Morimoto},\ and\ \citenamefont {Vishwanath}}]{Potter2016}%
  \BibitemOpen
  \bibfield  {author} {\bibinfo {author} {\bibfnamefont {A.~C.}\ \bibnamefont
  {Potter}}, \bibinfo {author} {\bibfnamefont {T.}~\bibnamefont {Morimoto}}, \
  and\ \bibinfo {author} {\bibfnamefont {A.}~\bibnamefont {Vishwanath}},\
  }\href@noop {} {\bibfield  {journal} {\bibinfo  {journal} {Phys. Rev. X}\
  }\textbf {\bibinfo {volume} {6}},\ \bibinfo {pages} {041001} (\bibinfo {year}
  {2016})}\BibitemShut {NoStop}%
\bibitem [{\citenamefont {{Nathan}}\ \emph {et~al.}()\citenamefont {{Nathan}},
  \citenamefont {{Rudner}}, \citenamefont {{Lindner}}, \citenamefont {{Berg}},\
  and\ \citenamefont {{Refael}}}]{Magnetization}%
  \BibitemOpen
  \bibfield  {author} {\bibinfo {author} {\bibfnamefont {F.}~\bibnamefont
  {{Nathan}}}, \bibinfo {author} {\bibfnamefont {M.~S.}\ \bibnamefont
  {{Rudner}}}, \bibinfo {author} {\bibfnamefont {N.~H.}\ \bibnamefont
  {{Lindner}}}, \bibinfo {author} {\bibfnamefont {E.}~\bibnamefont {{Berg}}}, \
  and\ \bibinfo {author} {\bibfnamefont {G.}~\bibnamefont {{Refael}}},\
  }\href@noop {} {\ }\Eprint {http://arxiv.org/abs/1610.03590}
  {arXiv:1610.03590} \BibitemShut {NoStop}%
\bibitem [{\citenamefont {Po}\ \emph {et~al.}(2016)\citenamefont {Po},
  \citenamefont {Fidkowski}, \citenamefont {Morimoto}, \citenamefont {Potter},\
  and\ \citenamefont {Vishwanath}}]{PhysRevX.6.041070}%
  \BibitemOpen
  \bibfield  {author} {\bibinfo {author} {\bibfnamefont {H.~C.}\ \bibnamefont
  {Po}}, \bibinfo {author} {\bibfnamefont {L.}~\bibnamefont {Fidkowski}},
  \bibinfo {author} {\bibfnamefont {T.}~\bibnamefont {Morimoto}}, \bibinfo
  {author} {\bibfnamefont {A.~C.}\ \bibnamefont {Potter}}, \ and\ \bibinfo
  {author} {\bibfnamefont {A.}~\bibnamefont {Vishwanath}},\ }\href {\doibase
  10.1103/PhysRevX.6.041070} {\bibfield  {journal} {\bibinfo  {journal} {Phys.
  Rev. X}\ }\textbf {\bibinfo {volume} {6}},\ \bibinfo {pages} {041070}
  (\bibinfo {year} {2016})}\BibitemShut {NoStop}%
\bibitem [{\citenamefont {{Bar Lev}}\ \emph {et~al.}(2015)\citenamefont {{Bar
  Lev}}, \citenamefont {Cohen},\ and\ \citenamefont {Reichman}}]{BarLev2015}%
  \BibitemOpen
  \bibfield  {author} {\bibinfo {author} {\bibfnamefont {Y.}~\bibnamefont {{Bar
  Lev}}}, \bibinfo {author} {\bibfnamefont {G.}~\bibnamefont {Cohen}}, \ and\
  \bibinfo {author} {\bibfnamefont {D.~R.}\ \bibnamefont {Reichman}},\
  }\href@noop {} {\bibfield  {journal} {\bibinfo  {journal} {Phys. Rev. Lett.}\
  }\textbf {\bibinfo {volume} {114}},\ \bibinfo {pages} {100601} (\bibinfo
  {year} {2015})}\BibitemShut {NoStop}%
\bibitem [{\citenamefont {Bera}\ \emph {et~al.}(2015)\citenamefont {Bera},
  \citenamefont {Schomerus}, \citenamefont {Heidrich-Meisner},\ and\
  \citenamefont {Bardarson}}]{Bera2015a}%
  \BibitemOpen
  \bibfield  {author} {\bibinfo {author} {\bibfnamefont {S.}~\bibnamefont
  {Bera}}, \bibinfo {author} {\bibfnamefont {H.}~\bibnamefont {Schomerus}},
  \bibinfo {author} {\bibfnamefont {F.}~\bibnamefont {Heidrich-Meisner}}, \
  and\ \bibinfo {author} {\bibfnamefont {J.~H.}\ \bibnamefont {Bardarson}},\
  }\href@noop {} {\bibfield  {journal} {\bibinfo  {journal} {Phys. Rev. Lett.}\
  }\textbf {\bibinfo {volume} {115}},\ \bibinfo {pages} {046603} (\bibinfo
  {year} {2015})}\BibitemShut {NoStop}%
\bibitem [{\citenamefont {Dunlap}\ and\ \citenamefont
  {Kenkre}(1986)}]{Dunlap1986}%
  \BibitemOpen
  \bibfield  {author} {\bibinfo {author} {\bibfnamefont {D.~H.}\ \bibnamefont
  {Dunlap}}\ and\ \bibinfo {author} {\bibfnamefont {V.~M.}\ \bibnamefont
  {Kenkre}},\ }\href@noop {} {\bibfield  {journal} {\bibinfo  {journal} {Phys.
  Rev. B}\ }\textbf {\bibinfo {volume} {34}},\ \bibinfo {pages} {3625}
  (\bibinfo {year} {1986})}\BibitemShut {NoStop}%
\bibitem [{\citenamefont {Grossmann}\ \emph {et~al.}(1991)\citenamefont
  {Grossmann}, \citenamefont {Dittrich}, \citenamefont {Jung},\ and\
  \citenamefont {H\"anggi}}]{Grossmann1991}%
  \BibitemOpen
  \bibfield  {author} {\bibinfo {author} {\bibfnamefont {F.}~\bibnamefont
  {Grossmann}}, \bibinfo {author} {\bibfnamefont {T.}~\bibnamefont {Dittrich}},
  \bibinfo {author} {\bibfnamefont {P.}~\bibnamefont {Jung}}, \ and\ \bibinfo
  {author} {\bibfnamefont {P.}~\bibnamefont {H\"anggi}},\ }\href@noop {}
  {\bibfield  {journal} {\bibinfo  {journal} {Phys. Rev. Lett.}\ }\textbf
  {\bibinfo {volume} {67}},\ \bibinfo {pages} {516} (\bibinfo {year}
  {1991})}\BibitemShut {NoStop}%
\bibitem [{\citenamefont {Lignier}\ \emph {et~al.}(2007)\citenamefont
  {Lignier}, \citenamefont {Sias}, \citenamefont {Ciampini}, \citenamefont
  {Singh}, \citenamefont {Zenesini}, \citenamefont {Morsch},\ and\
  \citenamefont {Arimondo}}]{Lignier2007}%
  \BibitemOpen
  \bibfield  {author} {\bibinfo {author} {\bibfnamefont {H.}~\bibnamefont
  {Lignier}}, \bibinfo {author} {\bibfnamefont {C.}~\bibnamefont {Sias}},
  \bibinfo {author} {\bibfnamefont {D.}~\bibnamefont {Ciampini}}, \bibinfo
  {author} {\bibfnamefont {Y.}~\bibnamefont {Singh}}, \bibinfo {author}
  {\bibfnamefont {A.}~\bibnamefont {Zenesini}}, \bibinfo {author}
  {\bibfnamefont {O.}~\bibnamefont {Morsch}}, \ and\ \bibinfo {author}
  {\bibfnamefont {E.}~\bibnamefont {Arimondo}},\ }\href@noop {} {\bibfield
  {journal} {\bibinfo  {journal} {Phys. Rev. Lett.}\ }\textbf {\bibinfo
  {volume} {99}},\ \bibinfo {pages} {220403} (\bibinfo {year}
  {2007})}\BibitemShut {NoStop}%
\bibitem [{\citenamefont {Eckardt}\ \emph {et~al.}(2009)\citenamefont
  {Eckardt}, \citenamefont {Holthaus}, \citenamefont {Lignier}, \citenamefont
  {Zenesini}, \citenamefont {Ciampini}, \citenamefont {Morsch},\ and\
  \citenamefont {Arimondo}}]{Eckardt2009}%
  \BibitemOpen
  \bibfield  {author} {\bibinfo {author} {\bibfnamefont {A.}~\bibnamefont
  {Eckardt}}, \bibinfo {author} {\bibfnamefont {M.}~\bibnamefont {Holthaus}},
  \bibinfo {author} {\bibfnamefont {H.}~\bibnamefont {Lignier}}, \bibinfo
  {author} {\bibfnamefont {A.}~\bibnamefont {Zenesini}}, \bibinfo {author}
  {\bibfnamefont {D.}~\bibnamefont {Ciampini}}, \bibinfo {author}
  {\bibfnamefont {O.}~\bibnamefont {Morsch}}, \ and\ \bibinfo {author}
  {\bibfnamefont {E.}~\bibnamefont {Arimondo}},\ }\href@noop {} {\bibfield
  {journal} {\bibinfo  {journal} {Phys. Rev. A}\ }\textbf {\bibinfo {volume}
  {79}},\ \bibinfo {pages} {013611} (\bibinfo {year} {2009})}\BibitemShut
  {NoStop}%
\bibitem [{\citenamefont {Eckardt}(2017)}]{Eckardt2016}%
  \BibitemOpen
  \bibfield  {author} {\bibinfo {author} {\bibfnamefont {A.}~\bibnamefont
  {Eckardt}},\ }\href {\doibase 10.1103/RevModPhys.89.011004} {\bibfield
  {journal} {\bibinfo  {journal} {Rev. Mod. Phys.}\ }\textbf {\bibinfo {volume}
  {89}},\ \bibinfo {pages} {011004} (\bibinfo {year} {2017})}\BibitemShut
  {NoStop}%
\bibitem [{\citenamefont {Eckardt}\ \emph {et~al.}(2005)\citenamefont
  {Eckardt}, \citenamefont {Weiss},\ and\ \citenamefont
  {Holthaus}}]{Eckardt2005}%
  \BibitemOpen
  \bibfield  {author} {\bibinfo {author} {\bibfnamefont {A.}~\bibnamefont
  {Eckardt}}, \bibinfo {author} {\bibfnamefont {C.}~\bibnamefont {Weiss}}, \
  and\ \bibinfo {author} {\bibfnamefont {M.}~\bibnamefont {Holthaus}},\
  }\href@noop {} {\bibfield  {journal} {\bibinfo  {journal} {Phys. Rev. Lett.}\
  }\textbf {\bibinfo {volume} {95}},\ \bibinfo {pages} {260404} (\bibinfo
  {year} {2005})}\BibitemShut {NoStop}%
\bibitem [{\citenamefont {Zenesini}\ \emph {et~al.}(2009)\citenamefont
  {Zenesini}, \citenamefont {Lignier}, \citenamefont {Ciampini}, \citenamefont
  {Morsch},\ and\ \citenamefont {Arimondo}}]{Zenesini2009}%
  \BibitemOpen
  \bibfield  {author} {\bibinfo {author} {\bibfnamefont {A.}~\bibnamefont
  {Zenesini}}, \bibinfo {author} {\bibfnamefont {H.}~\bibnamefont {Lignier}},
  \bibinfo {author} {\bibfnamefont {D.}~\bibnamefont {Ciampini}}, \bibinfo
  {author} {\bibfnamefont {O.}~\bibnamefont {Morsch}}, \ and\ \bibinfo {author}
  {\bibfnamefont {E.}~\bibnamefont {Arimondo}},\ }\href@noop {} {\bibfield
  {journal} {\bibinfo  {journal} {Phys. Rev. Lett.}\ }\textbf {\bibinfo
  {volume} {102}},\ \bibinfo {pages} {100403} (\bibinfo {year}
  {2009})}\BibitemShut {NoStop}%
\bibitem [{\citenamefont {Holthaus}\ \emph {et~al.}(1995)\citenamefont
  {Holthaus}, \citenamefont {Ristow},\ and\ \citenamefont
  {Hone}}]{Holthaus1995}%
  \BibitemOpen
  \bibfield  {author} {\bibinfo {author} {\bibfnamefont {M.}~\bibnamefont
  {Holthaus}}, \bibinfo {author} {\bibfnamefont {G.~H.}\ \bibnamefont
  {Ristow}}, \ and\ \bibinfo {author} {\bibfnamefont {D.~W.}\ \bibnamefont
  {Hone}},\ }\href@noop {} {\bibfield  {journal} {\bibinfo  {journal} {Phys.
  Rev. Lett.}\ }\textbf {\bibinfo {volume} {75}},\ \bibinfo {pages} {3914}
  (\bibinfo {year} {1995})}\BibitemShut {NoStop}%
\bibitem [{\citenamefont {Hone}\ and\ \citenamefont
  {Holthaus}(1993)}]{Hone1993}%
  \BibitemOpen
  \bibfield  {author} {\bibinfo {author} {\bibfnamefont {D.~W.}\ \bibnamefont
  {Hone}}\ and\ \bibinfo {author} {\bibfnamefont {M.}~\bibnamefont
  {Holthaus}},\ }\href@noop {} {\bibfield  {journal} {\bibinfo  {journal}
  {Phys. Rev. B}\ }\textbf {\bibinfo {volume} {48}},\ \bibinfo {pages} {15123}
  (\bibinfo {year} {1993})}\BibitemShut {NoStop}%
\bibitem [{\citenamefont {Martinez}\ and\ \citenamefont
  {Molina}(2006)}]{Martinez2006}%
  \BibitemOpen
  \bibfield  {author} {\bibinfo {author} {\bibfnamefont {D.~F.}\ \bibnamefont
  {Martinez}}\ and\ \bibinfo {author} {\bibfnamefont {R.~A.}\ \bibnamefont
  {Molina}},\ }\href@noop {} {\bibfield  {journal} {\bibinfo  {journal} {Phys.
  Rev. B}\ }\textbf {\bibinfo {volume} {73}},\ \bibinfo {pages} {073104}
  (\bibinfo {year} {2006})}\BibitemShut {NoStop}%
\bibitem [{\citenamefont {Drese}\ and\ \citenamefont
  {Holthaus}(1997)}]{Drese1997}%
  \BibitemOpen
  \bibfield  {author} {\bibinfo {author} {\bibfnamefont {K.}~\bibnamefont
  {Drese}}\ and\ \bibinfo {author} {\bibfnamefont {M.}~\bibnamefont
  {Holthaus}},\ }\href@noop {} {\bibfield  {journal} {\bibinfo  {journal}
  {Phys. Rev. Lett.}\ }\textbf {\bibinfo {volume} {78}},\ \bibinfo {pages}
  {2932} (\bibinfo {year} {1997})}\BibitemShut {NoStop}%
\bibitem [{\citenamefont {Roy}\ and\ \citenamefont {Das}(2015)}]{Das2015}%
  \BibitemOpen
  \bibfield  {author} {\bibinfo {author} {\bibfnamefont {A.}~\bibnamefont
  {Roy}}\ and\ \bibinfo {author} {\bibfnamefont {A.}~\bibnamefont {Das}},\
  }\href {\doibase 10.1103/PhysRevB.91.121106} {\bibfield  {journal} {\bibinfo
  {journal} {Phys. Rev. B}\ }\textbf {\bibinfo {volume} {91}},\ \bibinfo
  {pages} {121106} (\bibinfo {year} {2015})}\BibitemShut {NoStop}%
\bibitem [{\citenamefont {Bukov}\ \emph
  {et~al.}(2015{\natexlab{a}})\citenamefont {Bukov}, \citenamefont
  {D'Alessio},\ and\ \citenamefont {Polkovnikov}}]{Bukov2015}%
  \BibitemOpen
  \bibfield  {author} {\bibinfo {author} {\bibfnamefont {M.}~\bibnamefont
  {Bukov}}, \bibinfo {author} {\bibfnamefont {L.}~\bibnamefont {D'Alessio}}, \
  and\ \bibinfo {author} {\bibfnamefont {A.}~\bibnamefont {Polkovnikov}},\
  }\href@noop {} {\bibfield  {journal} {\bibinfo  {journal} {Advances in
  Physics}\ }\textbf {\bibinfo {volume} {64}},\ \bibinfo {pages} {139}
  (\bibinfo {year} {2015}{\natexlab{a}})}\BibitemShut {NoStop}%
\bibitem [{\citenamefont {Pal}\ and\ \citenamefont {Huse}(2010)}]{Pal2010}%
  \BibitemOpen
  \bibfield  {author} {\bibinfo {author} {\bibfnamefont {A.}~\bibnamefont
  {Pal}}\ and\ \bibinfo {author} {\bibfnamefont {D.~A.}\ \bibnamefont {Huse}},\
  }\href@noop {} {\bibfield  {journal} {\bibinfo  {journal} {Phys. Rev. B}\
  }\textbf {\bibinfo {volume} {82}},\ \bibinfo {pages} {174411} (\bibinfo
  {year} {2010})}\BibitemShut {NoStop}%
\bibitem [{\citenamefont {Feynman}\ \emph {et~al.}(1966)\citenamefont
  {Feynman}, \citenamefont {Leighton}, \citenamefont {Sands},\ and\
  \citenamefont {Lindsay}}]{feynman1966}%
  \BibitemOpen
  \bibfield  {author} {\bibinfo {author} {\bibfnamefont {R.~P.}\ \bibnamefont
  {Feynman}}, \bibinfo {author} {\bibfnamefont {R.~B.}\ \bibnamefont
  {Leighton}}, \bibinfo {author} {\bibfnamefont {M.}~\bibnamefont {Sands}}, \
  and\ \bibinfo {author} {\bibfnamefont {R.~B.}\ \bibnamefont {Lindsay}},\
  }\href {http://feynmanlectures.caltech.edu/} {\emph {\bibinfo {title} {The
  feynman lectures on physics, vol. 3: Quantum mechanics}}}\ (\bibinfo
  {publisher} {AIP},\ \bibinfo {year} {1966})\ Chap.~\bibinfo {chapter}
  {21}\BibitemShut {NoStop}%
\bibitem [{Note1()}]{Note1}%
  \BibitemOpen
  \bibinfo {note} {$J\DOTSB \mapstochar \rightarrow -J$ is a gauge symmetry of
  Eq. (\ref {eq: H0})}\BibitemShut {NoStop}%
\bibitem [{\citenamefont {Abanin}\ \emph {et~al.}(2015)\citenamefont {Abanin},
  \citenamefont {De~Roeck},\ and\ \citenamefont {Huveneers}}]{Abanin2015}%
  \BibitemOpen
  \bibfield  {author} {\bibinfo {author} {\bibfnamefont {D.~A.}\ \bibnamefont
  {Abanin}}, \bibinfo {author} {\bibfnamefont {W.}~\bibnamefont {De~Roeck}}, \
  and\ \bibinfo {author} {\bibfnamefont {F.~m.~c.}\ \bibnamefont {Huveneers}},\
  }\href {\doibase 10.1103/PhysRevLett.115.256803} {\bibfield  {journal}
  {\bibinfo  {journal} {Phys. Rev. Lett.}\ }\textbf {\bibinfo {volume} {115}},\
  \bibinfo {pages} {256803} (\bibinfo {year} {2015})}\BibitemShut {NoStop}%
\bibitem [{\citenamefont {Haldar}\ and\ \citenamefont {Das}(2017)}]{Das2017}%
  \BibitemOpen
  \bibfield  {author} {\bibinfo {author} {\bibfnamefont {A.}~\bibnamefont
  {Haldar}}\ and\ \bibinfo {author} {\bibfnamefont {A.}~\bibnamefont {Das}},\
  }\href {\doibase 10.1002/andp.201600333} {\bibfield  {journal} {\bibinfo
  {journal} {Annalen der Physik}\ }\textbf {\bibinfo {volume} {529}},\ \bibinfo
  {pages} {1600333} (\bibinfo {year} {2017})}\BibitemShut {NoStop}%
\bibitem [{\citenamefont {Serbyn}\ \emph {et~al.}(2013)\citenamefont {Serbyn},
  \citenamefont {Papi{\'{c}}},\ and\ \citenamefont {Abanin}}]{Serbyn2013}%
  \BibitemOpen
  \bibfield  {author} {\bibinfo {author} {\bibfnamefont {M.}~\bibnamefont
  {Serbyn}}, \bibinfo {author} {\bibfnamefont {Z.}~\bibnamefont {Papi{\'{c}}}},
  \ and\ \bibinfo {author} {\bibfnamefont {D.~A.}\ \bibnamefont {Abanin}},\
  }\href@noop {} {\bibfield  {journal} {\bibinfo  {journal} {Phys. Rev. Lett.}\
  }\textbf {\bibinfo {volume} {111}},\ \bibinfo {pages} {127201} (\bibinfo
  {year} {2013})}\BibitemShut {NoStop}%
\bibitem [{\citenamefont {Huse}\ \emph {et~al.}(2014)\citenamefont {Huse},
  \citenamefont {Nandkishore},\ and\ \citenamefont {Oganesyan}}]{Huse2014}%
  \BibitemOpen
  \bibfield  {author} {\bibinfo {author} {\bibfnamefont {D.~A.}\ \bibnamefont
  {Huse}}, \bibinfo {author} {\bibfnamefont {R.}~\bibnamefont {Nandkishore}}, \
  and\ \bibinfo {author} {\bibfnamefont {V.}~\bibnamefont {Oganesyan}},\
  }\href@noop {} {\bibfield  {journal} {\bibinfo  {journal} {Phys. Rev. B}\
  }\textbf {\bibinfo {volume} {90}},\ \bibinfo {pages} {174202} (\bibinfo
  {year} {2014})}\BibitemShut {NoStop}%
\bibitem [{\citenamefont {Ros}\ \emph {et~al.}(2015)\citenamefont {Ros},
  \citenamefont {Mueller},\ and\ \citenamefont {Scardicchio}}]{Ros2014}%
  \BibitemOpen
  \bibfield  {author} {\bibinfo {author} {\bibfnamefont {V.}~\bibnamefont
  {Ros}}, \bibinfo {author} {\bibfnamefont {M.}~\bibnamefont {Mueller}}, \ and\
  \bibinfo {author} {\bibfnamefont {A.}~\bibnamefont {Scardicchio}},\ }\href
  {\doibase http://dx.doi.org/10.1016/j.nuclphysb.2014.12.014} {\bibfield
  {journal} {\bibinfo  {journal} {Nuclear Physics B}\ }\textbf {\bibinfo
  {volume} {891}},\ \bibinfo {pages} {420 } (\bibinfo {year}
  {2015})}\BibitemShut {NoStop}%
\bibitem [{\citenamefont {Chandran}\ \emph {et~al.}(2015)\citenamefont
  {Chandran}, \citenamefont {Kim}, \citenamefont {Vidal},\ and\ \citenamefont
  {Abanin}}]{Chandran2015}%
  \BibitemOpen
  \bibfield  {author} {\bibinfo {author} {\bibfnamefont {A.}~\bibnamefont
  {Chandran}}, \bibinfo {author} {\bibfnamefont {I.~H.}\ \bibnamefont {Kim}},
  \bibinfo {author} {\bibfnamefont {G.}~\bibnamefont {Vidal}}, \ and\ \bibinfo
  {author} {\bibfnamefont {D.~A.}\ \bibnamefont {Abanin}},\ }\href@noop {}
  {\bibfield  {journal} {\bibinfo  {journal} {Phys. Rev. B}\ }\textbf {\bibinfo
  {volume} {91}},\ \bibinfo {pages} {085425} (\bibinfo {year}
  {2015})}\BibitemShut {NoStop}%
\bibitem [{Note2()}]{Note2}%
  \BibitemOpen
  \bibinfo {note} {Tuning $A/\omega $ away from a root of $\protect \mathcal
  {J}_0$ also decreases the time-dependent terms since $\DOTSB \sum@ \slimits@
  _{n\not =0}|\protect \mathcal {J}_n|^2=1-|\protect \mathcal {J}_0|^2$.
  However, since the localization length of $H_{eff}$ diverges as $J_{eff}$
  approaches $J_c$, we expect it to be the dominant factor in determining the
  shape of the phase diagram.}\BibitemShut {Stop}%
\bibitem [{\citenamefont {Zhang}\ \emph {et~al.}(2016)\citenamefont {Zhang},
  \citenamefont {Khemani},\ and\ \citenamefont {Huse}}]{Zhang2016}%
  \BibitemOpen
  \bibfield  {author} {\bibinfo {author} {\bibfnamefont {L.}~\bibnamefont
  {Zhang}}, \bibinfo {author} {\bibfnamefont {V.}~\bibnamefont {Khemani}}, \
  and\ \bibinfo {author} {\bibfnamefont {D.~A.}\ \bibnamefont {Huse}},\ }\href
  {\doibase 10.1103/PhysRevB.94.224202} {\bibfield  {journal} {\bibinfo
  {journal} {Phys. Rev. B}\ }\textbf {\bibinfo {volume} {94}},\ \bibinfo
  {pages} {224202} (\bibinfo {year} {2016})}\BibitemShut {NoStop}%
\bibitem [{\citenamefont {Srednicki}(1994)}]{SrednickiETH}%
  \BibitemOpen
  \bibfield  {author} {\bibinfo {author} {\bibfnamefont {M.}~\bibnamefont
  {Srednicki}},\ }\href {\doibase 10.1103/PhysRevE.50.888} {\bibfield
  {journal} {\bibinfo  {journal} {Phys. Rev. E}\ }\textbf {\bibinfo {volume}
  {50}},\ \bibinfo {pages} {888} (\bibinfo {year} {1994})}\BibitemShut
  {NoStop}%
\bibitem [{\citenamefont {Iyer}\ \emph {et~al.}(2013)\citenamefont {Iyer},
  \citenamefont {Oganesyan}, \citenamefont {Refael},\ and\ \citenamefont
  {Huse}}]{Iyer2013}%
  \BibitemOpen
  \bibfield  {author} {\bibinfo {author} {\bibfnamefont {S.}~\bibnamefont
  {Iyer}}, \bibinfo {author} {\bibfnamefont {V.}~\bibnamefont {Oganesyan}},
  \bibinfo {author} {\bibfnamefont {G.}~\bibnamefont {Refael}}, \ and\ \bibinfo
  {author} {\bibfnamefont {D.~A.}\ \bibnamefont {Huse}},\ }\href@noop {}
  {\bibfield  {journal} {\bibinfo  {journal} {Phys. Rev. B}\ }\textbf {\bibinfo
  {volume} {87}},\ \bibinfo {pages} {134202} (\bibinfo {year}
  {2013})}\BibitemShut {NoStop}%
\bibitem [{\citenamefont {Edwards}\ and\ \citenamefont
  {Anderson}(1975)}]{Edwards1975}%
  \BibitemOpen
  \bibfield  {author} {\bibinfo {author} {\bibfnamefont {S.~F.}\ \bibnamefont
  {Edwards}}\ and\ \bibinfo {author} {\bibfnamefont {P.~W.}\ \bibnamefont
  {Anderson}},\ }\href@noop {} {\bibfield  {journal} {\bibinfo  {journal}
  {Journal of Physics F: Metal Physics}\ }\textbf {\bibinfo {volume} {5}},\
  \bibinfo {pages} {965} (\bibinfo {year} {1975})}\BibitemShut {NoStop}%
\bibitem [{\citenamefont {Schreiber}\ \emph {et~al.}(2015)\citenamefont
  {Schreiber}, \citenamefont {Hodgman}, \citenamefont {Bordia}, \citenamefont
  {L{\"{u}}schen}, \citenamefont {Fischer}, \citenamefont {Vosk}, \citenamefont
  {Altman}, \citenamefont {Schneider},\ and\ \citenamefont
  {Bloch}}]{Schreiber2015}%
  \BibitemOpen
  \bibfield  {author} {\bibinfo {author} {\bibfnamefont {M.}~\bibnamefont
  {Schreiber}}, \bibinfo {author} {\bibfnamefont {S.~S.}\ \bibnamefont
  {Hodgman}}, \bibinfo {author} {\bibfnamefont {P.}~\bibnamefont {Bordia}},
  \bibinfo {author} {\bibfnamefont {H.~P.}\ \bibnamefont {L{\"{u}}schen}},
  \bibinfo {author} {\bibfnamefont {M.~H.}\ \bibnamefont {Fischer}}, \bibinfo
  {author} {\bibfnamefont {R.}~\bibnamefont {Vosk}}, \bibinfo {author}
  {\bibfnamefont {E.}~\bibnamefont {Altman}}, \bibinfo {author} {\bibfnamefont
  {U.}~\bibnamefont {Schneider}}, \ and\ \bibinfo {author} {\bibfnamefont
  {I.}~\bibnamefont {Bloch}},\ }\href@noop {} {\bibfield  {journal} {\bibinfo
  {journal} {Science}\ }\textbf {\bibinfo {volume} {349}},\ \bibinfo {pages}
  {842} (\bibinfo {year} {2015})}\BibitemShut {NoStop}%
\bibitem [{\citenamefont {Bordia}\ \emph {et~al.}(2017)\citenamefont {Bordia},
  \citenamefont {Luschen}, \citenamefont {Schneider}, \citenamefont {Knap},\
  and\ \citenamefont {Bloch}}]{Bordia2017}%
  \BibitemOpen
  \bibfield  {author} {\bibinfo {author} {\bibfnamefont {P.}~\bibnamefont
  {Bordia}}, \bibinfo {author} {\bibfnamefont {H.}~\bibnamefont {Luschen}},
  \bibinfo {author} {\bibfnamefont {U.}~\bibnamefont {Schneider}}, \bibinfo
  {author} {\bibfnamefont {M.}~\bibnamefont {Knap}}, \ and\ \bibinfo {author}
  {\bibfnamefont {I.}~\bibnamefont {Bloch}},\ }\href@noop {} {\bibfield
  {journal} {\bibinfo  {journal} {Nat. Phys.}\ }\textbf {\bibinfo {volume}
  {13}},\ \bibinfo {pages} {460} (\bibinfo {year} {2017})}\BibitemShut
  {NoStop}%
\bibitem [{\citenamefont {Luitz}\ \emph {et~al.}(2016)\citenamefont {Luitz},
  \citenamefont {Laflorencie},\ and\ \citenamefont {Alet}}]{Luitz2016a}%
  \BibitemOpen
  \bibfield  {author} {\bibinfo {author} {\bibfnamefont {D.~J.}\ \bibnamefont
  {Luitz}}, \bibinfo {author} {\bibfnamefont {N.}~\bibnamefont {Laflorencie}},
  \ and\ \bibinfo {author} {\bibfnamefont {F.}~\bibnamefont {Alet}},\
  }\href@noop {} {\bibfield  {journal} {\bibinfo  {journal} {Phys. Rev. B}\
  }\textbf {\bibinfo {volume} {93}},\ \bibinfo {pages} {060201} (\bibinfo
  {year} {2016})}\BibitemShut {NoStop}%
\bibitem [{\citenamefont {Bukov}\ \emph
  {et~al.}(2015{\natexlab{b}})\citenamefont {Bukov}, \citenamefont
  {Gopalakrishnan}, \citenamefont {Knap},\ and\ \citenamefont
  {Demler}}]{Bukov2015a}%
  \BibitemOpen
  \bibfield  {author} {\bibinfo {author} {\bibfnamefont {M.}~\bibnamefont
  {Bukov}}, \bibinfo {author} {\bibfnamefont {S.}~\bibnamefont
  {Gopalakrishnan}}, \bibinfo {author} {\bibfnamefont {M.}~\bibnamefont
  {Knap}}, \ and\ \bibinfo {author} {\bibfnamefont {E.}~\bibnamefont
  {Demler}},\ }\href@noop {} {\bibfield  {journal} {\bibinfo  {journal} {Phys.
  Rev. Lett.}\ }\textbf {\bibinfo {volume} {115}},\ \bibinfo {pages} {205301}
  (\bibinfo {year} {2015}{\natexlab{b}})}\BibitemShut {NoStop}%
\bibitem [{\citenamefont {Burin}()}]{Burin2017}%
  \BibitemOpen
  \bibfield  {author} {\bibinfo {author} {\bibfnamefont {A.~L.}\ \bibnamefont
  {Burin}},\ }\href@noop {} {\ }\Eprint {http://arxiv.org/abs/1702.01431}
  {arXiv:1702.01431} \BibitemShut {NoStop}%
\bibitem [{\citenamefont {Choi}\ \emph {et~al.}(2017)\citenamefont {Choi},
  \citenamefont {Abanin},\ and\ \citenamefont {Lukin}}]{Soonwon2017}%
  \BibitemOpen
  \bibfield  {author} {\bibinfo {author} {\bibfnamefont {S.}~\bibnamefont
  {Choi}}, \bibinfo {author} {\bibfnamefont {D.~A.}\ \bibnamefont {Abanin}}, \
  and\ \bibinfo {author} {\bibfnamefont {M.~D.}\ \bibnamefont {Lukin}},\
  }\href@noop {} {\  (\bibinfo {year} {2017})},\ \Eprint
  {http://arxiv.org/abs/arXiv:1703.03809} {arXiv:1703.03809} \BibitemShut
  {NoStop}%
  \bibitem [{\citenamefont {Dignam}\ and\ \citenamefont
  {de~Sterke}(2002)}]{Dignam2002}%
  \BibitemOpen
  \bibfield  {author} {\bibinfo {author} {\bibfnamefont {M.~M.}\ \bibnamefont
  {Dignam}}\ and\ \bibinfo {author} {\bibfnamefont {C.~M.}\ \bibnamefont
  {de~Sterke}},\ }\href {\doibase 10.1103/PhysRevLett.88.046806} {\bibfield
  {journal} {\bibinfo  {journal} {Phys. Rev. Lett.}\ }\textbf {\bibinfo
  {volume} {88}},\ \bibinfo {pages} {046806} (\bibinfo {year}
  {2002})}\BibitemShut {NoStop}%
\end{thebibliography}

\renewcommand{\theequation}{S\arabic{equation}}
\renewcommand{\thefigure}{S\arabic{figure}}
\setcounter{equation}{0}
\setcounter{figure}{0}

\clearpage
\pagebreak
\newpage

\section*{Supplemental Materials: Driving induced many-body localization}

\renewcommand{\theequation}{S\arabic{equation}}
\renewcommand{\thefigure}{S\arabic{figure}}

\maketitle

\subsection{Relaxation of initial product states at low frequencies}

In addition to the increase of the level statistics parameter with system size, the lack of localization at low driving frequencies is also manifested
in the relaxation of initial product states of particle occupations. When these states are driven at $\omega=3.5J$, their generalized
imbalance decays to a value which decreases with system size (Fig.
S1), as in the undriven case. Note that the decay as a function of
time is slower when compared to the decay in the undriven case. Presumably,
this is due to the proximity to the MBL transition \cite{Luitz2016a}
for this value of the driving frequency.

As the driving frequency
approaches the speculated critical frequency $\omega_{c}\approx4J$,
the remaining imbalance at long times declines much slower with system
size (Fig. S1 inset). The quick flattening of the slope in the inset
of Fig S1, for a small change of driving frequency, provides an independent
corroboration for the value of the critical frequency.

\begin{figure}[h!]
\begin{centering}
\includegraphics[width=8.6cm]{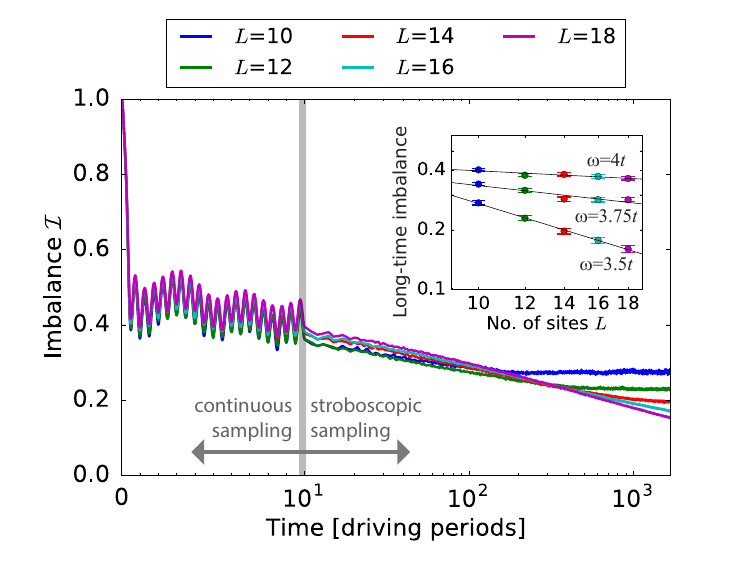}
\par\end{centering}
\protect\caption{Imbalance as a function of time for  $J_{eff}=0$ [$A/\omega=\left(A/\omega\right)^*$] at a low driving frequency $\omega=3.5J$. The imbalance at long times decays with system size, indicating that the system
is in the delocalized phase as expected from level statistics analysis. This is in contrast to the localized case for $\omega=5J$ (Fig. 3 in the main text), where the long-time imbalance does not depend on system size.
Inset: imbalance at long times (averaged over $1.4\times10^{3}T<t<1.5\times10^{3}T$)
as a function of system size for different driving frequencies. Error bars indicate one standard deviation over disorder realizations. The values
for the slopes are $-0.91\pm0.04$, $-0.33\pm0.12$, $-0.15\pm0.06$
for $\omega=3.5J$, $3.75J$, $4J$, respectively. The diminishing slope
as the driving frequency approaches $\omega=4J$ indicates slowing down of the dynamics, due to the proximity to the transition into the MBL phase. }
\end{figure}

\subsection{Phase boundaries between the ergodic and MBL phases}

The phase
boundaries in Fig. 1 were obtained by finite-size scaling of the quasienergy level statistics $\left\langle r\right\rangle$. Examples of such data
are given in Figs. 2, 4, S2. Namely, a consistent increase
of $\left\langle r\right\rangle$ with system size is interpreted as an indication
for the ergodic phase, while a consistent decrease of $\left\langle r\right\rangle$
with system size is interpreted as an indication for the MBL phase.
Error bars indicate parameter ranges where the trend in level statistics
with increasing system size is not statistically significant
(according to the error bar for $\left\langle r\right\rangle$, as shown for example in Figs. 2, 4, S2).

\begin{figure}[h!]
\begin{centering}
\includegraphics[width=8.6cm]{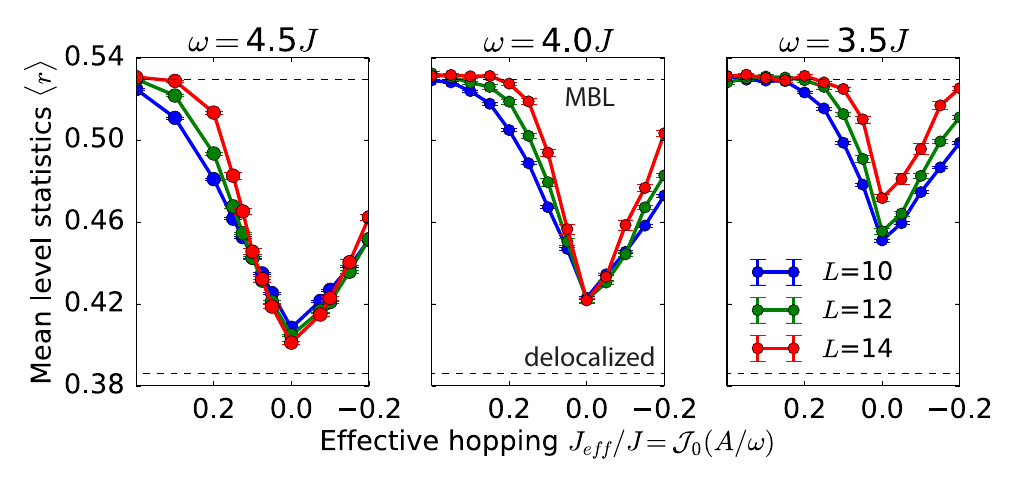}
\par\end{centering}
\protect\caption{Quasienergy level statistics as a function of driving amplitude at
a few driving frequencies near $\omega_c$. When the driving frequency is lowered,
the range of driving amplitudes which induce many body localization
narrows around $\left(A/\omega\right)^{*}$ corresponding to the first root of $\mathcal{J}_0$, for which $J_{eff}=0$. At $\omega=4J$, the
level statistics parameter at $\left(A/\omega\right)^{*}$ changes very slowly with system size, indicating
proximity to the critical frequency; at a lower frequency $\omega=3.5J$,
the level statistics parameter tends to the delocalized value at any
driving amplitude $A/\omega>0$ up to the first minimum of $\mathcal{J}_0$.}
\end{figure}

\subsection{Level statistics as a function of driving amplitude at low frequencies }

At $\omega=5J$
we found driving induced many-body localization in a range of driving
amplitudes around $\left(A/\omega\right)^*$, corresponding to the first root of $\mathcal{J}_0$, and for which $J_{eff}=0$ (Fig. 2 in main text). We expect this range to narrow for lower driving frequencies, and to vanish altogether for $\omega<\omega_{c}\approx4J$; at these frequencies, the drive fails to induce localization
even at $J_{eff}=0$ (Fig. 4). Indeed, when we perform finite-size
scaling of the level statistics as a function of driving amplitude
at frequencies lower than $5J$, we find that the range of localization-inducing
driving amplitudes continuously narrows, and all but shrinks to a
point $J_{eff}=0$ at $\omega=4J$ (Fig. S2 left, middle). Below this frequency,
the level statistics parameter increases with system size for any
driving amplitude $A/\omega>0$ up to the first minimum of $\mathcal{J}_0$ (Fig. S2 right), indicating that the drive fails to induce
localization for this range of driving amplitudes.

\subsection{Driving induced MBL beyond the first minimum of $\mathcal{J}_0$}
So far, we have analyzed a range of driving amplitudes $0\leq A/\omega\leq (A/\omega)_{\mathrm{min}}$, where $(A/\omega)_{\mathrm{min}}$ is the first minimum of the Bessel function (bright green in Fig. S3b).  We performed additional simulations which suggest a qualitatively similar phase diagram at higher driving amplitudes.

Specifically, we obtained the finite-size scaling of the quasienergy level statistics for driving amplitudes corresponding to the interval between the first minimum of $\mathcal{J}_0$ to its next maximum (dark green in Fig. S3b). The result of this analysis also shows localization in a range of driving amplitudes above a critical driving frequency. To determine the critical driving frequency for values of $A/\omega$ corresponding to $\left(A/\omega\right)^{**}$, the second root of $\mathcal{J}_0$, we varied the driving frequency while tuning the driving amplitude such that $A/\omega$ remains fixed. The results, shown in Fig S3a, indicate that the critical driving frequency for inducing MBL at $\left(A/\omega\right)^{**}$ is $\omega_c\approx3.25J$, which is lower compared the one found at the first root ($\omega_c\approx4J$).

\begin{figure}[b]
\begin{centering}
\includegraphics[width=8.6cm]{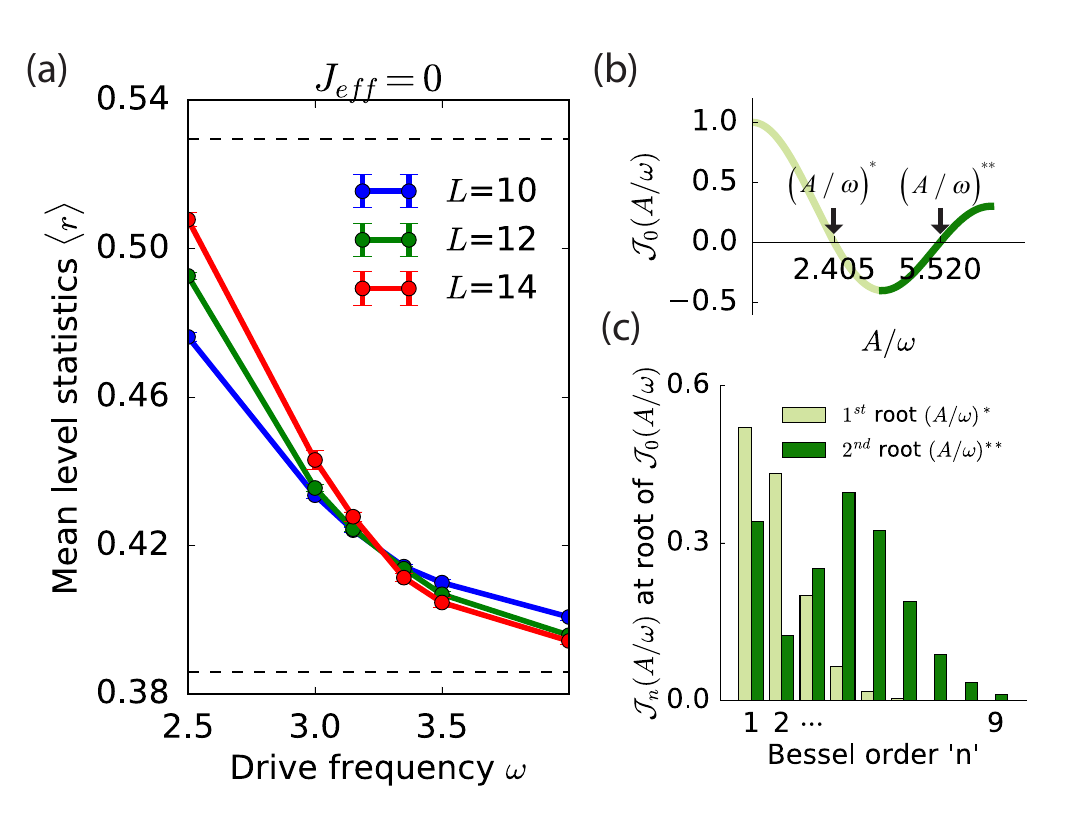}
\par\end{centering}
\protect\caption{Driving induced MBL at a driving amplitude corresponding to $\left(A/\omega\right)^{**}$, the second root of $\mathcal{J}_{0}\left(A/\omega\right)$. (a) Finite-size scaling of quasienergy level statistics as a function of driving frequency. The critical frequency at the second root ($\omega_c\approx3.25J$) is smaller than the one found at the first root, $\omega_c\approx4J$ . (b) Plot of the zeroth Bessel function $\mathcal{J}_0$: bright green indicates the amplitude range considered in the main text, additional simulations performed at the dark green amplitude range show qualitatively similar results. (c) Values of the various Bessel functions $\mathcal{J}_{n}$ at the first root $\left(A/\omega\right)^{*}$ (light green) vs. second root $\left(A/\omega\right)^{**}$ (dark green) of $\mathcal{J}_{0}\left(A/\omega\right)$, indicating the values of the non-zero Fourier modes of the Hamiltonian when the time-averaged Hamiltonian has no hopping term, $J_{eff}=0$. The values of the negative orders are related by $\mathcal{J}_{-n}\left(A/\omega\right)=(-1)^n\mathcal{J}_n\left(A/\omega\right)$.}
\end{figure}

This decrease in the critical driving frequency can be understood by comparing the Fourier spectrum of the Hamiltonian at the first two roots of the zeroth Bessel function (Fig. S3c). While the first Fourier mode of the Hamiltonian is the most dominant when $A/\omega$ is tuned to the first root of $\mathcal{J}_0$, at the second root of $\mathcal{J}_0$ the bulk of its Fourier spectrum shifts to the higher harmonics. Intuitively, driving at a larger amplitude therefore has a similar effect to increasing the driving frequency.

\subsection{Inducing MBL with a square-wave electric field}
\renewcommand{\thefigure}{S\arabic{figure}}
\begin{figure}[b!]
\begin{centering}
\includegraphics[width=8.6cm]{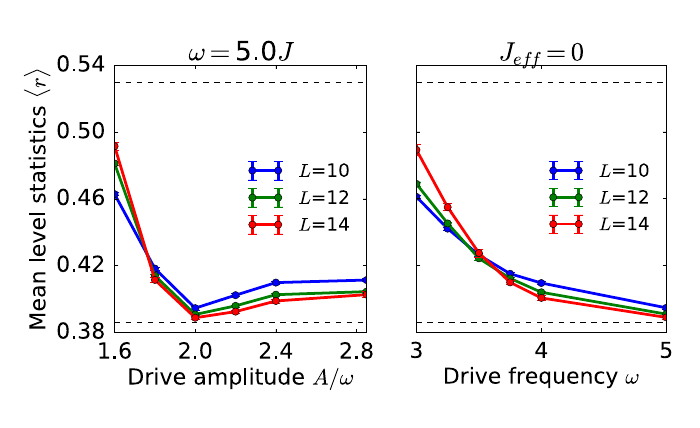}
\par\end{centering}
\protect\caption{Finite-size scaling of quasienergy level statistics $\left\langle r\right\rangle$ for the square-wave electric field (\ref{eq: S1}). Left: as a function of rescaled driving amplitude $A/\omega$ at a fixed driving frequency $\omega=5J$. Right: as a function of driving frequency for $A=2\omega$, corresponding to $J_{eff}=0$ for the square-wave drive [see Eq. (\ref{eq: Jeff_square_wave})] .}
\end{figure}

Our analysis so far focused on an AC electric field oscillating in time as $E(t)=A\cos(\omega t)$. However, the effective hopping in $H_{eff}$ is suppressed also for other functional forms for periodic time dependence of the AC electric field. We tested the possibility to induce an MBL phase with a square-wave AC electric field:
 \begin{equation}
\begin{aligned}
E\left(t\right)=\begin{cases}
A \quad \ \ 0 \leq t<T/2\\
-A \quad T/2 \leq t<T\
\end{cases}
\end{aligned}
\label{eq: S1}
\end{equation}

Such a field can be used for exact dynamical localization in models with hopping beyond nearest-neighbor \cite{Dignam2002,Eckardt2009}. In our model, performing the Peierls substitution and time-averaging the acquired phase leads to the effective hopping amplitude:
\begin{equation}
\begin{aligned}
J_{eff}/J=e^{-i\frac{\pi}{2}\frac{A}{\omega}}\cdot\sinc\left(\frac{\pi A}{2\omega}\right)
\end{aligned}
\label{eq: Jeff_square_wave}
\end{equation}
Again, finite-size scaling of quasienergy level statistics shows localization in a range of driving amplitudes around $J_{eff}=0$ (here obtained at $A/\omega=2$) below a critical driving frequency $\omega\approx3.5J$ (Fig. S4).

\subsection{Driving induced MBL at a lower filling fraction}

\renewcommand{\thefigure}{S\arabic{figure}}
\begin{figure}[b!]
\begin{centering}
\includegraphics[width=8.6cm]{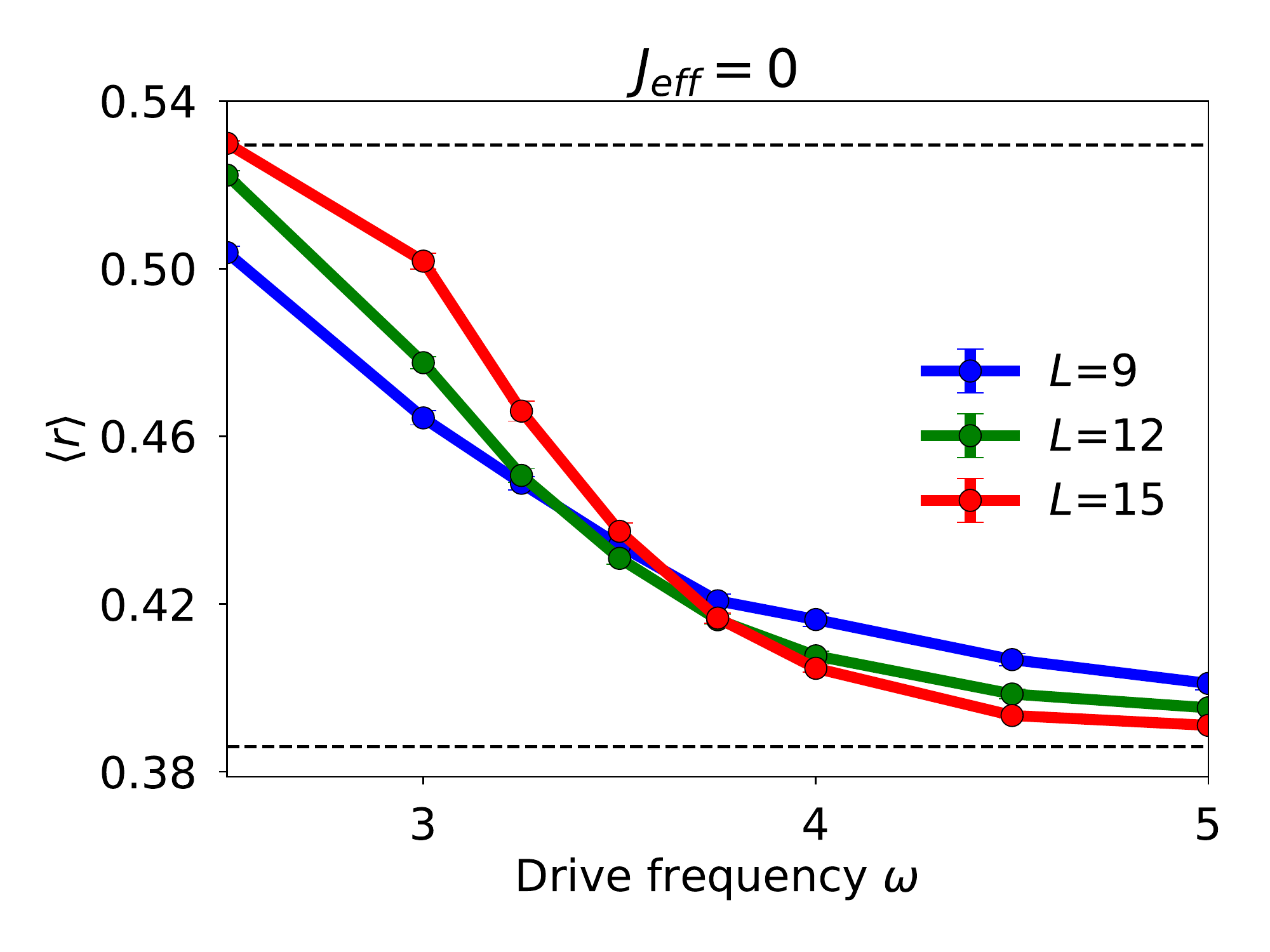}
\par\end{centering}
\protect\caption{Finite-size scaling of quasienergy level statistics $\left\langle r\right\rangle$ at filling fraction 1/3. We tune the driving frequency $\omega$ while fixing the rescaled driving amplitude $A/\omega$ at the first root of the zeroth Bessel function ($J_{eff}=0$). We find a critical frequency $\omega\approx3.75$ for inducing the MBL phase, which is slightly lower than the critical frequency found at half filling ($\omega\approx4$).}
\end{figure}
Our numerical simulations so far focused on the case of half filling. This filling fraction was chosen to maximize the width of the many-body spectrum for a given number of sites, thus minimizing finite-size effects.

We expect qualitatively similar results at different particle fillings. The critical frequency might slightly decrease though, since the interactions become effectively weaker away from half filling (this is apparent when the filling is decreased below 1/2, but is also true when it is increased due to particle-hole symmetry). In the parameter range we use for our simulations, weaker interactions imply stronger localization with a shorter localization length; therefore, the local spectrum becomes narrower and the critical frequency should correspondingly decrease.

To test this, we repeat the procedure of figure 4 at a different filling fraction $1/3$ (Fig. S5). Namely, we fix the rescaled driving amplitude $A/\omega$ at the first root of the zeroth Bessel function ($J_{eff}=0$) and perform finite-size scaling of the quasienergy level statistics as a function of the driving frequency $\omega$. Indeed, we find that the MBL phase is induced above a critical frequency $\omega\approx3.75$, which is slightly smaller than the critical frequency $\omega\approx4$ found at half filling.

\end{document}